# Security-Enhanced SC-FDMA Transmissions Using Temporal Artificial-Noise and Secret-Key Aided Schemes

Mohamed F. Marzban, *Student Member, IEEE,* Ahmed El Shafie, *Member, IEEE,* Naofal Al-Dhahir, *Fellow, IEEE,* Ridha Hamila, *Senior Member, IEEE*

**Abstract**

We investigate the physical layer security of uplink single-carrier frequency-division multiple-access (SC-FDMA) systems. Multiple users, Alices, send confidential messages to a common legitimate base-station, Bob, in the presence of an eavesdropper, Eve. To secure the legitimate transmissions, each user superimposes an artificial noise (AN) signal on the time-domain SC-FDMA data block. We reduce the computational and storage requirements at Bob's receiver by assuming simple per-sub-channel detectors. We assume that Eve has global channel knowledge of all links in addition to high computational capabilities, where she adopts high-complexity detectors such as single-user maximum likelihood (ML), multiuser minimum-mean-square-error (MMSE), and multiuser ML. We analyze the correlation properties of the time-domain AN signal and illustrate how Eve can exploit them to reduce the AN effects. We prove that the number of useful AN streams that can degrade Eves signal-to-noise ratio (SNR) is dependent on the *channel memories* of Alices-Bob and Alices-Eve links. Furthermore, we enhance the system security for the case of *partial* Alices-Bob channel knowledge at Eve, where Eve only knows the precoding matrices of the data and AN signals instead of knowing the entire Alices-Bob channel matrices, and propose a hybrid scheme that integrates temporal AN with channel-based secret-key extraction.

## I. INTRODUCTION

Wireless channels are vulnerable to eavesdropping security attacks due to their broadcast nature where an illegitimate node can overhear sensitive information about the communicating legitimate

Part of this paper will be presented at IEEE ICC 2018 [1].

This paper was made possible by NPRP grant number 8-627-2-260 from the Qatar National Research Fund (a member of Qatar Foundation). The statements made herein are solely the responsibility of the authors.

M. Marzban, A. El Shafie and N. Al-Dhahir are with the Department of Electrical and Computer Engineering, The University of Texas at Dallas, Tx, 75080, USA. (e-mail: mohamed.marzban@utdallas.edu, ahmed.salahelshafie@gmail.com, aldhahir@utdallas.edu). R. Hamila is with the Electrical Engineering at Qatar University (e-mail: hamila@qu.edu.qa).

nodes. Information security is conventionally preserved using encryption schemes implemented at the upper layers of the protocol stack. However, such schemes require high storage and computational capabilities. In addition, they preserve security under the assumption of limited computation capabilities and limited network parameters knowledge at the eavesdroppers. To enhance and complement the upper-layers security approaches, physical-layer (PHY) security was introduced to provide security at the waveform level by exploiting the time-varying random nature of the wireless channel.

In his seminal work, Wyner showed that secure transmissions are feasible as long as the wiretap channel, i.e., the transmitter-eavesdropper channel, is more degraded than the legitimate channel, i.e., the transmitter-legitimate receiver channel [2]. PHY security is quantified in terms of the secrecy capacity which is the maximum data rate with zero information leakage to the eavesdropper(s). Many research works investigated the impact of multiple transmit antennas at various nodes on the achievable secrecy rates by using data and artificial noise (AN) precoding schemes [3] where the transmissions are designed carefully to transmit the data signals in the direction of the legitimate channel vectors while the AN signals are transmitted along the directions orthogonal to the data vectors. The key idea of transmitting AN signals is to confuse the eavesdropping nodes and degrade their signal-plus-interference-to-noise ratios (SINRs) [3]–[5].

However, in Internet of Things (IoT) applications, where the communicating devices are equipped with limited processing and transmit power resources, only a single antenna is typically available at the wireless nodes. In addition, in uplink wireless transmissions, the limited transmit power and form factor constraints at portable devices typically prohibit equipping them with multiple transmit antennas/radio-frequency (RF) chains. In such practical scenarios, beamforming and spatial AN techniques cannot be implemented by the uplink transmitting nodes [3]–[5] which presents a major research challenge.

Another approach for PHY security is based on exploiting the randomness of the wireless channel for extracting identical secret-keys at the legitimate communicating nodes. Wireless channels are characterized by a reciprocity property at a given time/frequency/space resource. That is, two communicating nodes observe identical (or at least highly correlated) multi-path attributes at both ends of a wireless link at any instant of time. In a rich fading environment, channel variations over time maintain a source of randomness that can be exploited by the transmitter and the legitimate receiver to extract two identical sets of key symbols. Secret key extraction and generation from wireless channel measurements have been realized using different properties of the received signal, e.g., received signal strength (RSS) [6], phase differences [7], time delay (in wideband transmission) [8], [9], and channel state information (CSI) [10]. Since RSS is easy to measure in practice, it is often used in many scenarios (see, e.g., [6]). Upon agreeing on secure key symbols between the legitimate transmit-receiver pair, it can be used once as a one-time pad (OTP) to encrypt several data symbols. OTP encryption is perfectly secure and provably unbreakable as long as the number of secured data symbols is equal to the number of secret key symbols [11]. Unfortunately, the main restriction of OTP encryption is that the extracted number of secure key symbols from a random source (i.e., channel) is too low to encrypt all the transmitted data symbols from a transmitter [10], [12]–[14]. This is due to the fact that the channel reciprocity property can not be guaranteed unless the legitimate parties measure the channel simultaneously. Moreover, perfect independence between legitimate and eavesdropping links is not possible in some scenarios. Increasing the number of secret key symbols has been investigated in many works, e.g., [10], [12]–[14] and the references therein.

PHY security has been recently investigated for many transmission scenarios including those based on orthogonal-frequency division multiplexing (OFDM) modulation scheme and its multiple-access technology, orthogonal-frequency division-multiple access (OFDMA), see e.g. [15]–[21]. The key advantage of OFDM is its ability to convert a frequency-selective channel into a group of orthogonal flat-fading frequency sub-channels. In [15], the authors constrained their problem formulation to meet the secrecy rate requirements of all users in an OFDMA network while optimizing the energy harvested by all users. In [16], a key-generation scheme based on the precoding matrix indices was proposed for securing multiple-input multiple-output (MIMO)-OFDM systems. The authors of [18] investigated the PHY security of a single-input single-output single-antenna eavesdropper (SISOSE) OFDM system and proposed a temporal-AN injection scheme to increase the instantaneous secrecy rate (ISR). The problem formulation was then extended in [19] to investigate the PHY security of MIMOME-OFDM systems using a new hybrid spatial-temporal AN scheme. In [20] and [21], different AN precoding designs and power allocation schemes were investigated to enhance the ISR in OFDM systems.

Single-carrier frequency-division multiple access (SC-FDMA) has been adopted in the uplink of wireless standards such as the long term evolution (LTE) 4G standard [22]. In addition, it has recently been standardized as the multiple-access scheme for cellular vehicle-to-vehicle (C-V2V) and cellular-vehicle-to-everything (C-V2X) communication in LTE side-links [23]. SC-FDMA is based on single-carrier frequency-division equalization (SC-FDE) which evolved to realize the benefits of OFDM and single-carrier systems as discussed in, e.g., [24] and references therein. Unlike OFDMA systems, SC-FDMA has relatively low peak-to-average power ratio (PAPR) which makes it suitable for low-power devices. To capture frequency diversity in OFDMA, channel coding has to be performed across different sub-channels since decisions are made in the frequency domain. In contrast, SC-FDMA decodes the information in the time-domain which eliminates the need for coding across different sub-channels. Although securing uplink SC-FDMA transmissions is very critical, to the best of our knowledge, none of the previous research work has considered its PHY security (i.e., information-theoretic security).

Our main contributions in this paper are summarized as follows

- We design a secure scheme for SC-FDMA systems by adding a temporal (time-domain) AN signal to the data block. The key idea is to exploit the available temporal degrees of freedom due to the insertion of a cyclic prefix (CP) sequence to each data block. The AN precoding matrix is designed carefully to degrade the eavesdropper(s) channels only and to be canceled at the legitimate receiver.

- We allocate different power levels to the data and AN signals. To generalize the optimization setting, we assume a power fraction that determines the amount of power assigned to data signals relative to AN signals. Our AN design does not require the instantaneous channel state information (CSI) of Eve's link which makes our proposed scheme robust in mitigating eavesdropping attacks even when the eavesdroppers remain passive to conceal their presence.

- We investigate the average secrecy rate performance when Bob performs the conventional linear block ZF and MMSE detection strategies. We show that these strategies converge to simple per-sub-channel filtering. That is, the receiver design has low complexity at the legitimate receiver, which is a practical scenario for IoT devices.

- We consider the worst-case eavesdropping scenario where Eve is assumed to have very high computational capabilities and investigate three high-complexity detectors at Eve's receiver; namely, single-user ML, multiuser MMSE, and multiuser ML. We analyze the design complexity and the achieved data rate in each scenario. In addition, we consider the worst-case eavesdropping scenario where Eve has knowledge of the CSI of the Alices-Bob links and accounts for AN correlation in her detectors' designs. We compare the average secrecy rates under these detection strategies for different system design parameters.

- We show that the number of AN streams that Alice can inject without harming Bob is equal

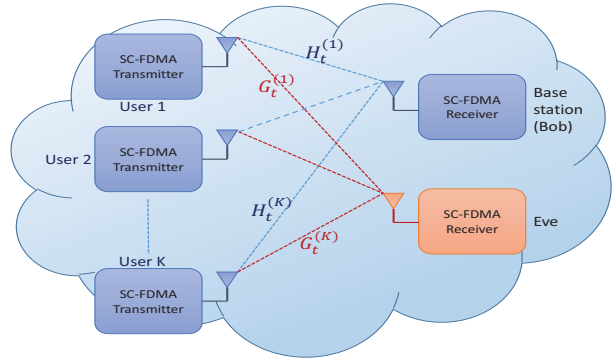

Fig. 1: Eavesdropping multiple-access uplink system model.

to the number of CP symbols. We prove that, out of those streams, the number of useful AN streams that can actually degrade Eve's SINR is the maximum of the Alice-Bob's and Alice-Eve's channel memories. The remaining AN streams lie in the null space of Alice-Eve's channel matrix. This prevents Alice from wasting power on useless AN streams.

- We derive a new lower bound on the average secrecy rate and show that, at high input SNR, the average secrecy rate is a linear function of Alice's transmit power level (in dB scale). Hence, unlike the case of no AN injection where the ISR becomes independent of Alice's transmit power level and saturates with it, in our investigated AN-aided scheme, the transmit power can still increase the ISR. Moreover, for a large number of data symbols per SC-FDMA block, the achievable average secrecy rate is a linear function of the number of useful AN streams.

- To enhance the system security for the case of partial Alices-Bob channel knowledge at Eve, we propose a hybrid scheme that integrates temporal AN with channel-based secret-key extraction. In our proposed scheme, we exploit the channel variations to extract secret key symbols which are used by Alices to encrypt an equivalent number of data symbols using an OTP. The encrypted data symbols are then multiplexed with the remaining unencrypted data symbols in the time domain. Finally, temporal-AN signals are added to the entire SC-FDMA block to secure the unencrypted data symbols. The encrypted data symbols are information-theoretically secured and their generation guarantees a positive total ISR. In addition, they provide an additional source of interference (in addition to temporal-AN symbols) that degrades Eve's reception of the unencrypted data symbols.

*Notation:* Lower- and upper-case bold letters denote vectors and matrices, respectively, while the subscripts, $(\cdot)_\text{f}$ and $(\cdot)_\text{t}$, refer to frequency-domain and time-domain quantities, respectively. $\mathbf{I}_N$ and $\mathbf{F}_N$ denote, respectively, the $N \times N$ identity matrix and the fast Fourier transform (FFT) matrix. $\mathbb{C}^{M \times N}$ and $\mathbb{R}^{M \times N}$ denote the set of all $M \times N$ complex and real matrices, respectively. $(\cdot)^\top$ and $(\cdot)^*$ denote the transpose and Hermitian (i.e., complex-conjugate transpose) operations, respectively, and $\mathbb{E}\{\cdot\}$, denotes statistical expectation. $\mathbf{A}^{(k)}$ denotes the matrix $\mathbf{A}$ associated with user $k$ and $[\mathbf{A}]_{i,1:N}$ is the $i$-th row of the matrix $\mathbf{A} \in \mathbb{C}^{M \times N}$. $\mathbf{0}_{M \times N}$ denotes the all-zero matrix with size $M \times N$, and $|\cdot|$ denotes the absolute value. $\text{diag}\{\cdot\}$ denotes a diagonal matrix whose diagonal elements are the enclosed entries, while $\text{diag}\{\cdot\}_{i,i}$ represents the $i$-th diagonal element of the enclosed matrix. $[\cdot]^+ = \max\{0, \cdot\}$ returns the maximum between the argument and zero, and $[\mathbf{A} \; \mathbf{B}]$ represents the horizontal concatenation of matrices $\mathbf{A}$ and $\mathbf{B}$. $\mathcal{CN}(0, \boldsymbol{\Sigma})$ denotes the complex Gaussian distribution with zero mean and covariance matrix denoted by $\boldsymbol{\Sigma}$. Throughout this paper, the term, *symbol*, is used to denote a single quadrature amplitude modulation (QAM) constellation symbol while the entire SC-FDMA symbol is denoted as, *block*.

## II. SYSTEM MODEL

Consider an SC-FDMA uplink communication system consisting of $K$ legitimate users, called Alices, and a single common base-station, called Bob. Each user sends confidential information messages to Bob in the presence of a passive eavesdropper, called Eve, who overhears the ongoing communications as depicted in Fig. 1. Let $M$ denote the total number of sub-channels of the SC-FDMA block and $N \leq M$ is the number of sub-channels dedicated to each user. We assume an equal sub-channel allocation to all users, i.e., $N = M/K$. All nodes are assumed to be equipped with a single antenna. SC-FDMA transmissions can be considered as FFT-precoded OFDMA transmissions where the symbols are precoded using an $N$-point FFT before applying an OFDMA modulator at the transmitter side. Let $\mathbf{x}_\text{t}^{(k)}$ denote the $N \times 1$ zero-mean unit-variance data symbols transmitted by user $k \in \{1, 2, \ldots, K\}$. Following the 3GPP-LTE standard, we assume that each Alice allocates her transmission power equally across her data symbols to maintain the low PAPR advantage of the SC-FDMA system. Let $p_\text{x}^{(k)}$ denote the transmission power of the $k$-th user's symbols. The data symbols vector is first transformed into a frequency-domain symbols vector, denoted by $\mathbf{x}_\text{f}^{(k)}$, using an $N$-point FFT as $\mathbf{x}_\text{f}^{(k)} = \mathbf{F}_N \mathbf{x}_\text{t}^{(k)}$.

We follow the LTE uplink localized FDMA sub-channel mapping strategy [23] where each user is assigned an adjacent set of sub-channels. Let $\mathbf{S}^{(k)}$ denote the $M \times N$ binary sub-channel

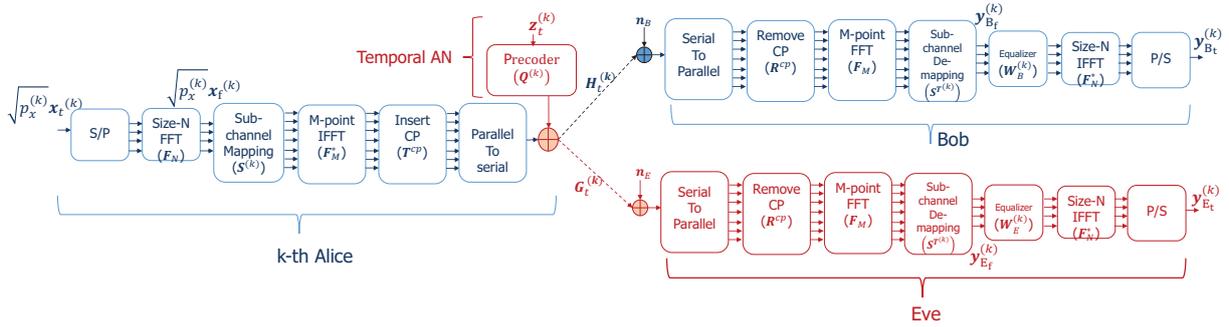

Fig. 2: System model for SC-FDMA with temporal AN.

mapping matrix which assigns $N$ out of the total $M$ sub-channels to user $k$ and is given by

$$\left[\mathbf{S}^{(k)}\right]_{i,j} = \begin{cases} 1, & \text{sub-channel } i \text{ is assigned to user } k \text{ and mapped to element } j \\ 0, & \text{Otherwise} \end{cases} \quad (1)$$

Next, an $M$-point IFFT is used to transform the signal back to the time domain. Prior to transmission, a CP sequence of size $M_{\text{cp}}$ is inserted at the beginning of each SC-FDMA data block using a CP insertion matrix, denoted by $\mathbf{T}^{\text{cp}} = \left[\mathbf{E}^{\top}_{M_{\text{cp}} \times M} \; \mathbf{I}_M\right]^{\top} \in \mathbb{R}^{(M+M_{\text{cp}}) \times M}$ with $\mathbf{E} = \left[\mathbf{0}_{M_{\text{cp}} \times (M-M_{\text{cp}})} \; \mathbf{I}_{M_{\text{cp}}}\right]$.

Let $\text{A}_k$, B, and E, denote the $k$-th Alice, Bob, and Eve, respectively. Let $L_{\text{B}}^{(k)}$ and $L_{\text{E}}^{(k)}$ denote the channel memories of the $\text{A}_k - \text{B}$ and $\text{A}_k - \text{E}$ links, respectively. The CP is designed to be longer than the channel memories of all links to eliminate the inter-block interference. We assume a quasi-static block-fading channel model where the channel impulse response (CIR) coefficients are assumed to remain constant during a single block and vary independently from one block to another. Let $\mathbf{H}_{\text{t}}^{(k)} \in \mathbb{C}^{(M+M_{cp}) \times (M+M_{cp})}$ and $\mathbf{G}_{\text{t}}^{(k)} \in \mathbb{C}^{(M+M_{cp}) \times (M+M_{cp})}$ denote the Toeplitz time-domain channel matrices of the $\text{A}_k - \text{B}$ and $\text{A}_k - \text{E}$ links, respectively, which have the following form for the $\text{A}_k - \text{B}$ link

$$\mathbf{H}_{\text{t}}^{(k)} = \begin{bmatrix} h^{(k)}(0) & 0 & 0 & \ldots & 0 \\ h^{(k)}(1) & h^{(k)}(0) & 0 & \ldots & 0 \\ \vdots & \ddots & \ddots & \ldots & \vdots \\ h^{(k)}(L_{\text{B}}^{(k)}) & \ddots & \ddots & \ldots & 0 \\ \vdots & \ddots & \ddots & \ldots & 0 \\ 0 & \ldots & h^{(k)}(L_{\text{B}}^{(k)}) & \ldots & h^{(k)}(0) \end{bmatrix} \quad (2)$$

where $h^{(k)}(i)$ represents the $i$-th CIR tap for the $\text{A}_k-\text{B}$ link. Each CIR tap of the $\text{A}_k-\text{B}$ and $\text{A}_k-\text{E}$ links has an average gain of $\sigma^{2(k)}_{\text{A}-\text{B}}$ and $\sigma^{2(k)}_{\text{A}-\text{E}}$, respectively.

At the receiver side, the CP is first removed using a CP removal matrix, denoted by $\mathbf{R}^{\text{cp}} = \begin{bmatrix} \mathbf{0}_{M \times M_{\text{cp}}} \mathbf{I}_N \end{bmatrix} \in \mathbb{R}^{M \times (M+M_{\text{cp}})}$. After that, an $M$-point FFT operation and sub-channel de-mapping operation, denoted by $\mathbf{S}^{(k)\top}$, are applied to extract the $k$-th user's data symbols in the frequency-domain as follows

$$\mathbf{y}_{\text{Bf}}^{(k)} = \mathbf{S}^{(k)\top} \mathbf{F}_M \mathbf{R}^{\text{cp}} \mathbf{H}_{\text{t}}^{(k)} \mathbf{T}^{\text{cp}} \mathbf{F}_M^* \mathbf{S}^{(k)} \mathbf{F}_N \sqrt{p_{\text{x}}^{(k)}} \mathbf{x}_{\text{t}}^{(k)} + \mathbf{n}_{\text{B}}^{(k)} = \sqrt{p_{\text{x}}^{(k)}} \mathbf{H}_{\text{f}}^{(k)} \mathbf{x}_{\text{f}}^{(k)} + \mathbf{n}_{\text{B}}^{(k)} \quad (3)$$

$$\mathbf{y}_{\text{Ef}}^{(k)} = \mathbf{S}^{(k)\top} \mathbf{F}_M \mathbf{R}^{\text{cp}} \mathbf{G}_{\text{t}}^{(k)} \mathbf{T}^{\text{cp}} \mathbf{F}_M^* \mathbf{S}^{(k)} \mathbf{F}_N \sqrt{p_{\text{x}}^{(k)}} \mathbf{x}_{\text{t}}^{(k)} + \mathbf{n}_{\text{E}}^{(k)} = \sqrt{p_{\text{x}}^{(k)}} \mathbf{G}_{\text{f}}^{(k)} \mathbf{x}_{\text{f}}^{(k)} + \mathbf{n}_{\text{E}}^{(k)} \quad (4)$$

where $\mathbf{n}_{\text{B}}^{(k)} \sim \mathcal{CN}(0, \kappa_{\text{B}})$ and $\mathbf{n}_{\text{E}}^{(k)} \sim \mathcal{CN}(0, \kappa_{\text{E}})$ denote the complex $N \times 1$ zero-mean circularly-symmetric additive white Gaussian noise (AWGN) vectors at Bob and Eve, respectively, and $\mathbf{H}_{\text{f}}^{(k)} = \mathbf{S}^{(k)\top} \mathbf{F}_M \mathbf{R}^{\text{cp}} \mathbf{H}_{\text{t}}^{(k)} \mathbf{T}^{\text{cp}} \mathbf{F}_M^* \mathbf{S}^{(k)} \in \mathbb{C}^{N \times N}$ and $\mathbf{G}_{\text{f}}^{(k)} = \mathbf{S}^{(k)\top} \mathbf{F}_M \mathbf{R}^{\text{cp}} \mathbf{G}_{\text{t}}^{(k)} \mathbf{T}^{\text{cp}} \mathbf{F}_M^* \mathbf{S}^{(k)} \in \mathbb{C}^{N \times N}$ denote the diagonal frequency-domain channel matrices of the $\text{A}_k-\text{B}$ and $\text{A}_k-\text{E}$ links, respectively.

In the post-processing stage, channel equalization is carried out in the frequency domain. Then, an $N$-point IFFT operation is performed to recover the time-domain data symbols. The $k$-th Alice's $N \times 1$ time-domain linearly equalized received vectors at Bob and Eve are given by

$$\mathbf{y}_{\text{Bt}}^{(k)} = \mathbf{F}_N^* \mathbf{W}_{\text{B}}^{(k)} \mathbf{H}_{\text{f}}^{(k)} \mathbf{F}_N \sqrt{p_{\text{x}}^{(k)}} \mathbf{x}_{\text{t}}^{(k)} + \mathbf{F}_N^* \mathbf{W}_{\text{B}}^{(k)} \mathbf{n}_{\text{B}}^{(k)} \quad (5)$$

$$\mathbf{y}_{\text{Et}}^{(k)} = \mathbf{F}_N^* \mathbf{W}_{\text{E}}^{(k)} \mathbf{G}_{\text{f}}^{(k)} \mathbf{F}_N \sqrt{p_{\text{x}}^{(k)}} \mathbf{x}_{\text{t}}^{(k)} + \mathbf{F}_N^* \mathbf{W}_{\text{E}}^{(k)} \mathbf{n}_{\text{E}}^{(k)} \quad (6)$$

where $\mathbf{W}_{\text{B}}^{(k)}$ and $\mathbf{W}_{\text{E}}^{(k)} \in \mathbb{C}^{N \times N}$ are the linear equalization matrix filters for the $k$-th Alice data at Bob and Eve, respectively. We will investigate different design choices for $\mathbf{W}_{\text{B}}^{(k)}$ and $\mathbf{W}_{\text{E}}^{(k)}$ in addition to non-linear equalization at Eve in later sections. The SC-FDMA system block diagram is depicted in Fig. 2.

## III. Temporal AN Design in SC-FDMA

To confuse Eve, Alice sacrifices a portion of her transmit power to inject an AN signal in the time domain to exploit the temporal degrees of freedom provided by the CP. A precoding matrix, denoted by $\mathbf{Q}^{(k)} \in \mathbb{C}^{(M+M_{\text{cp}}) \times M_{\text{cp}}}$, projects the AN vector to span the null space of the channel matrix between the $k$-th Alice and Bob as follows

$$\mathbf{R}^{\text{cp}} \mathbf{H}_{\text{t}}^{(k)} \mathbf{Q}^{(k)} = 0 \quad (7)$$

Given that $\mathbf{Q}^{(k)}$ has $M_{\text{cp}}$ orthonormal basis vectors, the $k$-th Alice can transmit a maximum of $M_{\text{cp}}$ AN streams without harming Bob. Let $\mathbf{z}_{\text{t}}^{(k)} \sim \mathcal{CN}(0, \boldsymbol{\Sigma}_z^{(k)})$ denote the $M_{\text{cp}} \times 1$ AN symbols transmitted by the $k$-th Alice to confuse Eve. To cover all possible eavesdroping directions and maximize the AN randomness at Eve, each Alice sends uncorrelated AN symbols. Hence, the AN covariance matrix, $\boldsymbol{\Sigma}_z^{(k)}$ is diagonal with the diagonal entries equal to the AN power for each AN stream. Let $p_{\text{t}}$ denote the total transmission power available at each user. We assume that the $k$-th Alice divides her power such that a fraction, $\alpha^{(k)} < 1$ of her total transmit power is allocated to data transmission, i.e. $p_{\text{x}} = \alpha^{(k)} \frac{p_{\text{t}}}{N}$, and the remaining fraction, $(1 - \alpha^{(k)})$, is allocated to AN. Each Alice inserts the AN signal in the null space of her channel matrix with Bob and, hence, the received signal at Bob in (5) will not change for all users. In contrast, the $k$-th Alice signal vector received at Eve in (6) is impaired with AN signals produced by all $K$ Alices as follows

$$\mathbf{y}_{\text{Et}}^{(k)} = \mathbf{F}_N^* \mathbf{W}_{\text{E}}^{(k)} \left( \mathbf{G}_{\text{f}}^{(k)} \mathbf{F}_N \sqrt{p_{\text{x}}^{(k)}} \mathbf{x}_{\text{t}}^{(k)} + \mathbf{n}_{\text{E}}^{(k)} + \mathbf{S}^{(k)\top} \sum_{j=1}^{K} \mathbf{F}_M \mathbf{R}^{\text{cp}} \mathbf{G}_{\text{t}}^{(j)} \mathbf{Q}^{(j)} \mathbf{z}_{\text{t}}^{(j)} \right) \qquad (8)$$

We emphasize here that this AN design does not assume knowledge of the Alices-Eve channels at the legitimate nodes. This is the best-case scenario for Eve where she remains passive and does not transmit to hide her CSI from the legitimate nodes. Furthermore, we assume that Eve has full knowledge of her own channel with all users as well as full knowledge of the Alices-Bob channels.

## IV. Linear SC-FDMA Data Detection with Temporal AN

In this section, we investigate two linear-detection strategies for SC-FDMA and derive the ISR for each strategy.

### A. Zero-Forcing (ZF) Detector

A simple detection strategy at Bob and Eve is the zero-forcing (ZF) detector which completely eliminates the channel distortion by inverting the diagonal frequency domain channel matrices, i.e., $\mathbf{W}_{\text{B}}^{(k)} = \mathbf{H}_{\text{f}}^{(k)-1}$ and $\mathbf{W}_{\text{E}}^{(k)} = \mathbf{G}_{\text{f}}^{(k)-1}$ at Bob and Eve, respectively. Hence, the frequency-domain ZF-equalized received signals at Bob and Eve can be expressed, respectively, as

$$\mathbf{y}_{\text{Bt}}^{(k)} = \sqrt{p_{\text{x}}^{(k)}} \mathbf{x}_{\text{t}}^{(k)} + \mathbf{F}_N^* \mathbf{H}_{\text{f}}^{(k)-1} \mathbf{n}_{\text{B}}^{(k)} \qquad (9)$$

$$\mathbf{y}_{\text{Et}}^{(k)} = \sqrt{p_{\text{x}}^{(k)}} \mathbf{x}_{\text{t}}^{(k)} + \mathbf{F}_N^* \mathbf{G}_{\text{f}}^{(k)-1} \left( \mathbf{n}_{\text{E}}^{(k)} + \mathbf{S}^{(k)\top} \sum_{j=1}^{K} \mathbf{O}^{(j)} \mathbf{z}_{\text{t}}^{(j)} \right) \qquad (10)$$

where the term, $\mathbf{S}^{(k)\top}\mathbf{O}^{(j)}\mathbf{z}_{\mathrm{t}}^{(j)}$, represents the interference signal affecting the $k$-th user data at Eve's receiver due to the AN transmission from user $j$, and $\mathbf{O}^{(j)} = \mathbf{F}_M\mathbf{R}^{\mathrm{cp}}\mathbf{G}_{\mathrm{t}}^{(j)}\mathbf{Q}^{(j)}$. At Bob, ZF equalization correlates the AWGN symbols resulting in a noise covariance matrix of $\mathbb{E}\left(\mathbf{F}_N^*\mathbf{H}_{\mathrm{f}}^{(k)-1}\mathbf{n}_{\mathrm{B}}^{(k)}\mathbf{n}_{\mathrm{B}}^{(k)*}\mathbf{H}_{\mathrm{f}}^{(k)-1*}\mathbf{F}_N\right) = \kappa_{\mathrm{B}}\mathbf{F}_N^*|\mathbf{H}_{\mathrm{f}}^{(k)-1}|^2\mathbf{F}_N$ where $\kappa_{\mathrm{B}}$ is the AWGN power at Bob before equalization. At Eve, ZF equalization does not only correlate the AWGN, but it also ignores the AN which can severely impact ZF performance.

After ZF equalization, each data symbol is decoded independently. Hence, the data rates of the $\mathrm{A}_k - \mathrm{B}$ and $\mathrm{A}_k - \mathrm{E}$ links are, respectively, given by

$$R_{\mathrm{B}}^{(k)} = \sum_{i=1}^{N} \log_2\left(1 + \frac{p_{x_i}^{(k)}}{\gamma_{\mathrm{B},n_i}^{2(k)}}\right) \tag{11}$$

$$R_{\mathrm{E}}^{(k)} = \sum_{i=1}^{N} \log_2\left(1 + \frac{p_{x_i}^{(k)}}{\gamma_{\mathrm{E},n_i}^{2(k)} + \gamma_{\mathrm{E},z_i}^{2(k)}}\right) \tag{12}$$

where $\gamma_{\mathrm{B},n_i}^{2(k)}$ and $\gamma_{\mathrm{E},n_i}^{2(k)}$ represent the AWGN powers across the $i$-th symbol at Bob and Eve, respectively, which are given by

$$\gamma_{\mathrm{B},n_i}^{2(k)} = \mathrm{diag}\left\{\mathbf{F}_N^*|\mathbf{H}_{\mathrm{f}}^{(k)-1}|^2\mathbf{F}_N\right\}_{i,i}\kappa_{\mathrm{B}} \tag{13}$$

$$\gamma_{\mathrm{E},n_i}^{2(k)} = \mathrm{diag}\left\{\mathbf{F}_N^*|\mathbf{G}_{\mathrm{f}}^{(k)-1}|^2\mathbf{F}_N\right\}_{i,i}\kappa_{\mathrm{E}} \tag{14}$$

where $\kappa_{\mathrm{E}}$ denotes the noise power before ZF equalization at Eve and $\gamma_{\mathrm{E},z_i}^{2(k)}$ represents the AN power affecting the $i$-th data symbol at Eve's receiver which is given by

$$\gamma_{\mathrm{E},z_i}^{(k)2} = \mathrm{diag}\left\{\mathbf{F}_N^*\mathbf{G}_{\mathrm{f}}^{(k)-1}\mathbf{S}^{(k)\top}\left(\sum_{j=1}^{K}\mathbf{O}^{(j)}\mathbf{\Sigma}_{\mathrm{z}}^{(j)}\mathbf{O}^{(j)*}\right) \times \mathbf{S}^{(k)}\mathbf{G}_{\mathrm{f}}^{(k)-1*}\mathbf{F}_N\right\}_{i,i} \tag{15}$$

Using the achievable rates of the $\mathrm{A}_K - \mathrm{B}$ and $\mathrm{A}_K - \mathrm{E}$ links, the ISR of user $k$ (in bits/sec/Hz) is given by

$$R_s^{(k)} = \frac{1}{M + M_{\mathrm{cp}}}\left[R_{\mathrm{B}}^{(k)} - R_{\mathrm{E}}^{(k)}\right]^+ \tag{16}$$

The sum ISR of all users is thus given by

$$R_s = \sum_k R_s^{(k)} \tag{17}$$

**Remark 1.** *From (10) and (12), we observe that all the $M$ data symbols at Eve are perturbed by the AN interference from every Alice even though each Alice transmits only $M_{\mathrm{cp}} \ll M$ AN streams. This suggests that the AN interference is highly correlated across the $M$ data symbols* *within the same SC-FDMA data block. The ZF strategy detects each symbol independently and does not account for this AN correlation. In the next subsections, we show that Eve can exploit the AN correlation to reduce its effects by adopting either an MMSE or an ML detection strategy. In particular, the MMSE detector jointly filters all symbols within an SC-FDMA block simultaneously but detects each symbol separately using a hard-decision device (slicer). The ML detector enables Eve to further exploit the AN correlation and minimize the interference by jointly filtering and detecting all symbols within an SC-FDMA block in the time domain.*

### B. Linear Block Minimum Mean-Square Error (MMSE) Detector

The MMSE detector [25] tackles the well-known noise enhancement problem of the ZF detector. The price paid is the need to estimate the noise covariance matrix. In this subsection, we investigate the achieved ISR when Bob and Eve use the block linear MMSE detection strategy which balances between minimizing inter-symbol interference (ISI) and noise enhancement. For the MMSE detector, the receive matrix filter, denoted by $\mathbf{W}$, can be expressed as follows

$$\mathbf{W} = \mathbf{R}_{\mathrm{xy}}\mathbf{R}_{\mathrm{yy}}^{-1} \tag{18}$$

where $\mathbf{R}_{\mathrm{xy}}$ is the cross-covariance matrix between the frequency-domain transmitted data vector, $\mathbf{x}_{\mathrm{f}}$, and the frequency-domain received signal vector, $\mathbf{y}_{\mathrm{f}}$, and $\mathbf{R}_{\mathrm{yy}}$ is the received signal covariance matrix in frequency-domain. Hence, the error-covariance matrix is given by

$$\mathbf{R}_{\mathrm{ee}} = \mathbf{R}_{\mathrm{xx}} - \mathbf{R}_{\mathrm{xy}}\mathbf{R}_{\mathrm{yy}}^{-1}\mathbf{R}_{\mathrm{yx}} \tag{19}$$

where $\mathbf{R}_{\mathrm{xx}}$ is the covariance matrix of the frequency-domain transmitted data vector. Since Bob experiences an AWGN noise along with the ISI, his linear MMSE detection filter for the $k$-th user's symbols is given by

$$\mathbf{W}_{\mathrm{B}}^{(k)} = p_x^{(k)}\mathbf{H}_{\mathrm{f}}^{(k)^*}\left(p_x^{(k)}\mathbf{H}_{\mathrm{f}}^{(k)}\mathbf{H}_{\mathrm{f}}^{(k)^*} + \kappa_{\mathrm{B}}\mathbf{I}_N\right)^{-1} \tag{20}$$

Since $\mathbf{H}_{\mathrm{f}}^{(k)}$ is a diagonal matrix, Bob can filter each sub-channel independently where the $i$-th sub-channel filter is then given by

$$w_{\mathrm{B_i}}^{(k)} = \frac{\left[\mathbf{H}_{\mathrm{f}}^{(k)}\right]_{i,i}^*}{|\mathbf{H}_{\mathrm{f}\ i,i}^{(k)}|^2 + \frac{\kappa_{\mathrm{B}}}{p_x^{(k)}}} \tag{21}$$

Bob's estimation error covariance matrix is given by

$$\mathbf{R}_{\mathrm{ee,B}}^{(k)} = \left(\mathbf{I}_N + \frac{p_x^{(k)}}{\kappa_{\mathrm{B}}}\mathbf{H}_{\mathrm{f}}^{(k)^*}\mathbf{H}_{\mathrm{f}}^{(k)}\right)^{-1} \tag{22}$$

where $\left[\mathbf{H}_\text{f}^{(k)}\right]_{i,i}$ is the $i$-th diagonal entry of $\mathbf{H}_\text{f}^{(k)}$. The inverse in (22) is easy to compute since the inverted matrix is diagonal. Therefore, the $i$-th sub-channel estimation error variance is given by

$$\left[\mathbf{R}_{\text{ee,B}}^{(k)}\right]_{i,i} = \frac{1}{1 + \frac{p_\text{x}^{(k)}}{\kappa_\text{B}}|\mathbf{H}_{\text{f}\ i,i}^{(k)}|^2} \tag{23}$$

The unbiased decision point SINR for symbol $i$ can be expressed as

$$\text{SINR}_{\text{B}_i}^{(k)} = \frac{1}{\left[\mathbf{R}_{\text{ee,B}}^{(k)}\right]_{i,i}} - 1 \tag{24}$$

Hence, the rate of the $k$-th Alice-Bob link is given by

$$R_\text{B}^{(k)} = \sum_{i=1}^{N} \log_2\left(1 + \text{SINR}_{\text{B}_i}^{(k)}\right) = \sum_{i=1}^{N} \log_2\left(1 + \frac{p_\text{x}^{(k)}|\mathbf{H}_{\text{f}\ i,i}^{(k)}|^2}{\kappa_\text{B}}\right) \tag{25}$$

In addition to ISI and AWGN, Eve's received signal is perturbed by the temporal AN interference transmitted by all $K$ users. Let $\tilde{\mathbf{n}}_\text{E}^{(k)}$ denote the $N \times 1$ equivalent noise vector which includes both the AN and the AWGN, i.e.,

$$\tilde{\mathbf{n}}_\text{E}^{(k)} = \mathbf{n}_\text{E}^{(k)} + \mathbf{S}^{(k)\top}\sum_{j=1}^{K}\mathbf{O}^{(j)}\mathbf{z}_\text{t}^{(j)} \tag{26}$$

We consider the worst-case security scenario where Eve has perfect knowledge of the CSI of the Alice-Bob link. Hence, she knows the null space precoder and can exploit the AN correlation across the data symbols within the same SC-FDMA block in her MMSE filter design to mitigate the AN effect. The cross-covariance matrix between the frequency-domain received and the transmitted data vectors, denoted by $\mathbf{R}_{\text{yx,E}}^{(k)}$, as well as the received signal covariance matrix at Eve's receiver, denoted by $\mathbf{R}_{\text{yy,E}}^{(k)}$, are given, respectively, by

$$\mathbf{R}_{\text{yx,E}}^{(k)} = \sqrt{p_\text{x}^{(k)}}\mathbf{G}_\text{f}^{(k)} \tag{27}$$

$$\mathbf{R}_{\text{yy,E}}^{(k)} = p_\text{x}^{(k)}\mathbf{G}_\text{f}^{(k)}\mathbf{G}_\text{f}^{(k)*} + \mathbf{R}_{\tilde{\mathbf{n}}\tilde{\mathbf{n}},\text{E}}^{(k)} \tag{28}$$

where $\mathbf{R}_{\tilde{\mathbf{n}}\tilde{\mathbf{n}},\text{E}}^{(k)}$ is the AN-plus-AWGN covariance matrix at Eve's receiver which is given by

$$\mathbf{R}_{\tilde{\mathbf{n}}\tilde{\mathbf{n}},\text{E}}^{(k)} = \kappa_\text{E}\mathbf{I}_N + \mathbf{S}^{(k)\top}\left(\sum_{j=1}^{K}\mathbf{O}^{(j)}\boldsymbol{\Sigma}_\text{z}^{(j)}\mathbf{O}^{(j)*}\right)\mathbf{S}^{(k)} \tag{29}$$

From (19) and (28), the MMSE filter matrix and the error-covariance matrix at Eve's receiver are given, respectively, by

$$\mathbf{W}_\text{E}^{(k)} = \sqrt{p_\text{x}^{(k)}}\mathbf{G}_\text{f}^{(k)*}\left(p_\text{x}^{(k)}\mathbf{G}_\text{f}^{(k)}\mathbf{G}_\text{f}^{(k)*} + \mathbf{R}_{\tilde{\mathbf{n}}\tilde{\mathbf{n}},\text{E}}^{(k)}\right)^{-1} \tag{30}$$

$$\mathbf{R}_{\text{ee,E}}^{(k)} = \mathbf{I}_N - p_{\text{x}}^{(k)} \mathbf{G}_{\text{f}}^{(k)*} \left( p_{\text{x}}^{(k)} \mathbf{G}_{\text{f}}^{(k)} \mathbf{G}_{\text{f}}^{(k)*} + \mathbf{R}_{\tilde{n}\tilde{n},\text{E}}^{(k)} \right)^{-1} \mathbf{G}_{\text{f}}^{(k)} \quad (31)$$

The decision-point SINR at the $i$-th sub-channel can be obtained from (31) and (24) as follows

$$\text{SINR}_{\text{E}_i}^{(k)} = \frac{1}{\left[\mathbf{R}_{\text{ee,E}}^{(k)}\right]_{i,i}} - 1 \quad (32)$$

Hence, the data rate of the $k$-th Alice-Eve link is given by

$$\mathbf{R}_{\text{E}}^{(k)} = \sum_{i=1}^{N} \log_2 \left(1 + \text{SINR}_{\text{E}_i}^{(k)}\right) \quad (33)$$

The ISR and the sum ISR are computed, respectively, from (16) and (17). We emphasize that, from (21) and (23), Bob's block MMSE detector simplifies to per-sub-channel filtering. In contrast, from (30) and (31), Eve's block MMSE detector is much more complicated since it requires matrix inversion and joint filtering of all the received symbols within the SC-FDMA block.

In the following two sections, we derive the achieved ISR when even more complicated detectors are implemented at Eve while Bob is still constrained to the simple per-sub-channel detectors of Section IV.

## V. Single-user Maximum Likelihood (ML) Detector at Eve

With equally-likely input symbols, the minimum error rate single-user detection strategy is the joint ML detection of a user's all data symbols. This can be realized by performing exhaustive search over all of the $k$-th user's data symbols within the SC-FDMA transmission block which requires very high computational complexity and can not be implemented using a linear filter. Using ML detection, Eve can further exploit the AN correlation across the $k$-th user's data symbols and reduce the AN effects. In this case, the data rate of the $k$-th Alice-Eve link is given by

$$R_{\text{E}}^{(k)} = \log_2 \det \left( \mathbf{I}_N + p_{\text{x}}^{(k)} \mathbf{G}_{\text{f}}^{(k)} \mathbf{G}_{\text{f}}^{(k)*} \left\{ \kappa_{\text{E}} \mathbf{I}_N + \mathbf{S}^{(k)\top} \left( \sum_{j=1}^{\text{K}} \mathbf{O}^{(j)} \mathbf{\Sigma}_{\text{z}}^{(j)} \mathbf{O}^{(j)*} \right) \mathbf{S}^{(k)} \right\}^{-1} \right) \quad (34)$$

Similar to the linear detection methods, the ISR and the sum ISR under single-user ML detection can be derived, respectively, using (16) and (17). Even though ML is the optimal detector in the sense that it minimizes the error rate for equally-likely information symbols, its complexity increases exponentially with the number of users' symbols per block.

**Proposition 1.** *Let $L_\mathrm{u}^{(k)}$ denote the maximum of the channel memories of the $k$-th Alice-Bob and $k$-th Alice-Eve links (i.e., $L_\mathrm{u}^{(k)} = \max(L_\mathrm{B}^{(k)}, L_\mathrm{E}^{(k)})$). There are $(M_\mathrm{cp} - L_\mathrm{u}^{(k)})$ useless AN directions that project the AN signals in the null space of the equivalent channel matrix of the $k$-th Alice-Eve link, and $L_\mathrm{u}^{(k)}$ useful AN directions that can degrade Eve's instantaneous rate. The useful AN streams can be extracted by designing the null space precoder as in (54).*

*Proof.* See Appendix A. □

This proposition suggests that although the $k$-th Alice can transmit a maximum of $M_\mathrm{cp}$ AN streams in Bob's null space, only $L_\mathrm{u}^{(k)} \leq M_\mathrm{cp}$ of them are useful in the sense that they hurt Eve. Hence, increasing $M_\mathrm{cp}$ without changing the channel memories of the Alices-Bob and Alices-Eve links will not increase the ISR.

**Proposition 2.** *Assuming single-user ML detection at Eve, at very high input SNR levels, the average secrecy rate (in bits/sec/Hz) is lower bounded as follows*

$$\frac{1}{M + M_\mathrm{cp}} \mathbb{E}\left\{\left[R_\mathrm{B}^{(k)} - R_\mathrm{E}^{(k)}\right]^+\right\} \gtrapprox \frac{\max_k\{L_\mathrm{u}^{(k)}\}}{M + M_\mathrm{cp}} \log_2\left(\frac{p_\mathrm{t}}{N\kappa_\mathrm{B}}\right) \tag{35}$$

*Proof.* See Appendix B. □

This is a very promising result since it shows that using our proposed temporal AN design, increasing the input power level, $p_\mathrm{t}$, can increase the average secrecy rate unlike the no-AN case. Moreover, the average secrecy rate is always positive and increases linearly with the number of useful AN streams, $L_\mathrm{u}^{(k)}$. This result also reveals that, as long as the power fraction parameter, $\alpha^{(k)}$, is not equal to zero or one, the average secrecy rate is independent of $\alpha^{(k)}$.

## VI. Multiuser Detection at Eve

In the previous sections, the single-user detectors at Bob and Eve separated each user's data using the sub-channel de-mapping matrix, $\mathbf{S}^{(k)^\top}$, prior to data detection. However, in multiuser detection, multiple users' data symbols are detected simultaneously. From performance perspective, multiuser detection is equivalent to single-user detection when the data streams of the different users do not interfere with each other. This is the case at Bob's receiver since there is no AN. However, at Eve's receiver, each temporal AN symbol interferes with all users' data symbols within the size-$M$ SC-FDMA block. This suggests that the AN is correlated across all users' symbols within the SC-FDMA block as discussed in Remark 1. Hence, multi-user detection can enhance the detection performance at Eve at the expense of increased complexity. In the following sub-sections, we will assume that Eve can perform multiuser detection and we will evaluate the achieved sum ISR.

## A. Multiuser MMSE Detector

To perform multiuser detection, Eve re-arranges all users' frequency-domain data symbols into a single $M \times 1$ vector denoted by, $\mathbf{y}_{\mathrm{Ef}} = \left[ \mathbf{y}_{\mathrm{Ef}}^{(1)}, \mathbf{y}_{\mathrm{Ef}}^{(2)}, \ldots \mathbf{y}_{\mathrm{Ef}}^{(K)} \right]^\top$, which is expressed as follows

$$\mathbf{y}_{\mathrm{Ef}} = \mathbf{G}_{\mathrm{f}} \mathbf{P}_{\mathrm{x}}^{\frac{1}{2}} \mathbf{x}_{\mathrm{f}} + \tilde{\mathbf{n}}_{\mathrm{E}} \tag{36}$$

The $M \times M$ diagonal matrix $\mathbf{G}_{\mathrm{f}}$ represents the equivalent channel matrix at Eve for all users in which the $k$-th block matrix is $\mathbf{G}_{\mathrm{f}}^{(k)}$, $\mathbf{x}_{\mathrm{f}}$ denotes the $M \times 1$ frequency-domain transmitted data vector of all users which is given by $\left[ \mathbf{x}_{\mathrm{f}}^{(1)}, \mathbf{x}_{\mathrm{f}}^{(2)}, \ldots, \mathbf{x}_{\mathrm{f}}^{(K)} \right]^\top$. In addition, $\tilde{\mathbf{n}}_{\mathrm{E}}$ denotes the equivalent AWGN-plus-AN vector at Eve, which is given by $\left[ \tilde{\mathbf{n}}_{\mathrm{E}}^{(1)}, \tilde{\mathbf{n}}_{\mathrm{E}}^{(2)}, \ldots, \tilde{\mathbf{n}}_{\mathrm{E}}^{(K)} \right]^\top$, and the diagonal matrix $\mathbf{P}_{\mathrm{x}} \in \mathbb{R}^{M \times M}$ represents the data transmit power matrix which contains the transmission power levels for all users' symbols and is given by

$$\mathbf{P}_{\mathrm{x}} = \mathrm{diag}\left( \underbrace{p_{\mathrm{x}}^{(1)}, p_{\mathrm{x}}^{(1)}, \ldots p_{\mathrm{x}}^{(1)}}_{N \text{ terms}}, \underbrace{p_{\mathrm{x}}^{(2)}, \ldots p_{\mathrm{x}}^{(2)}}_{N \text{ terms}}, \ldots, \underbrace{p_{\mathrm{x}}^{(K)}, \ldots p_{\mathrm{x}}^{(K)}}_{N \text{ terms}} \right) \tag{37}$$

Similar to (29), (30) and (31), the $M \times M$ noise covariance matrix, size-$M$ MMSE filter matrix, and the estimation error covariance matrix can be expressed, respectively, as

$$\begin{aligned} \mathbf{R}_{\tilde{\mathbf{n}}\tilde{\mathbf{n}},\mathrm{E}} &= \kappa_{\mathrm{E}} \mathbf{I}_M + \mathbf{G}_{\mathrm{f}}^{-1} \left( \sum_{j=1}^{K} \mathbf{O}^{(j)} \mathbf{\Sigma}_{\mathrm{z}}^{(j)} \mathbf{O}^{(j)*} \right) \mathbf{G}_{\mathrm{f}}^{-1*} \\ \mathbf{W}_{\mathrm{E}} &= \mathbf{P}_{\mathrm{x}}^{\frac{1}{2}} \mathbf{G}_{\mathrm{f}}^{*} \left( \mathbf{G}_{\mathrm{f}} \mathbf{P}_{\mathrm{x}} \mathbf{G}_{\mathrm{f}}^{*} + \mathbf{R}_{\tilde{\mathbf{n}}\tilde{\mathbf{n}},\mathrm{E}}^{(k)} \right)^{-1} \\ \mathbf{R}_{\mathrm{ee},\mathrm{E}} &= \mathbf{I}_M - \mathbf{P}_{\mathrm{x}}^{\frac{1}{2}} \mathbf{G}_{\mathrm{f}}^{*} \left( \mathbf{G}_{\mathrm{f}} \mathbf{P}_{\mathrm{x}} \mathbf{G}_{\mathrm{f}}^{*} + \mathbf{R}_{\tilde{\mathbf{n}}\tilde{\mathbf{n}},\mathrm{E}}^{(k)} \right)^{-1} \mathbf{G}_{\mathrm{f}} \mathbf{P}_{\mathrm{x}}^{\frac{1}{2}} \end{aligned} \tag{38}$$

We emphasize here that, in contrast to single-user MMSE detection, the multiuser MMSE detector at Eve exploits the temporal AN correlation across all $M$-sub-channels of all users to mitigate its effects. However, this requires much higher computational complexity at Eve since the $N \times N$ matrix inversion in the single-user detection is now an $M \times M$ matrix inversion as shown in (38). The decision-point SINR for the $i$-th symbol is given by

$$\mathrm{SINR}_{\mathrm{E}_i} = \frac{1}{[\mathbf{R}_{\mathrm{ee},\mathrm{E}}]_{i,i}} - 1 \tag{39}$$

Hence, the rate of the $k$-th Alice-Eve link is given by

$$\mathbf{R}_{\mathrm{E}}^{(k)} = \sum_{i \in \nu^{(k)}} \log_2\left(1 + \mathrm{SINR}_{\mathrm{E}_i}\right) \qquad (40)$$

where $\nu^{(k)}$ are the set of symbols assigned to user $k$. Similarly, the ISR and the sum ISR can be computed as in (16) and (17), respectively.

## B. Multiuser ML Detector

Similar to the MMSE case, Eve can jointly decode all users data simultaneously using an ML detector. This is the optimal detection strategy for Eve in terms of minimizing the error rate assuming equally-likely data symbols. However, it requires massive computation complexity. At Eve's receiver, the data symbols are first re-arranged as in (36). Then, Eve performs exhaustive search across the $M$ data symbols of all users within the SC-FDMA block. In this case, Eve's sum-rate is given by

$$\sum_k R_{\mathrm{E}}^{(k)} = \log_2 \det\left(\mathbf{I}_M + \mathbf{G}_{\mathrm{f}}\mathbf{P}_{\mathrm{x}}\mathbf{G}_{\mathrm{f}}^*\left(\kappa_{\mathrm{E}}\mathbf{I}_M + \sum_{j=1}^{K} \mathbf{O}^{(j)}\boldsymbol{\Sigma}_{\mathrm{z}}^{(j)}\mathbf{O}^{(j)*}\right)^{-1}\right) \qquad (41)$$

Therefore, the sum ISR is given by

$$R_{\mathrm{s}} = \frac{1}{M + M_{\mathrm{cp}}}\left[\sum_k R_{\mathrm{B}}^{(k)} - \sum_k R_{\mathrm{E}}^{(k)}\right]^+ \qquad (42)$$

## VII. HYBRID TEMPORAL AN SECRET KEY SCHEME

In the propsed temporal AN scheme, the $k$-th Alice and Bob exchange control and training sequences to estimate their links channels. Bob designs the AN precoding matrices using Eqn. (7) based on the channel estimates and only feedbacks the designed AN precoder to the $k$-th Alice. In this section, for more practical considerations, we relax the global channel knowledge assumption at Eve. We assume that Eve exploits the shared training sequences to estimate her own channel with all Alices and Bob ($\mathrm{A}_k - \mathrm{E}$ and $\mathrm{B} - \mathrm{E}$ channels) perfectly. In addition, she overhears the shared AN precoding matrices but does not know the exact channel matrices between Alices and Bob ($\mathrm{A}_k - \mathrm{B}$ channels). In this case, we propose enhancing the system security using a hybrid temporal-AN/secret-key (TAN-SK) scheme.

Next, we describe our proposed hybrid scheme. The $k$-th Alice ($\mathrm{A}_k$) selects her data symbols from a set of constellation points, denoted by $\mathcal{S}$, that is approximated to follow a Gaussian probability distribution [26]. Then, $\mathrm{A}_k$ and Bob start exchanging training and control signals to estimate the CSI of their links. Based on the channel reciprocity property, $A_k$ and Bob exploit their channel estimates to generate identical secret key symbols using any of the approaches in, e.g., [10], [14] and the references therein[1]. Let $\mathbf{r}^{(k)}$ denote the generated common secret key of size $N_{\text{en}}^{(k)}$ where $\mathbf{r}^{(k)} = [r_1^{(k)}, r_2^{(k)}, \cdots, r_{N_{\text{en}}^{(k)}}^{(k)}]$ with $r_i^{(k)} \in \mathcal{S}$. Eve is assumed to be sufficiently distant from all Alices and Bob such that the channel responses of her links $A_k - E$ and $B - E$ are independent from those of $A_k - B$ links and, hence, she can not generate the same secret key symbols. Upon agreeing on secure key symbols, $A_k$ uses the generated secret key symbols to encrypt an equivalent number of the data symbols through an OTP encryption scheme. The OTP encryption scheme provides a one-to-one mapping between the time-domain unencrypted data symbol, $x_{\text{t}_i}^{(k)}$, and its encrypted version, $x_{\text{t,en}_i}^{(k)}$, given the knowledge of the secret key, $r_i^{(k)}$. The security is unbreakable since each secret key symbol, $r_i^{(k)}$, is uniformly distributed over $\mathcal{S}$. Hence, no information is leaked to any eavesdropping node that intercepts the encrypted data symbols [26].

In general, the key generation rates are lower than the achievable data rates. Only a fraction of the data symbols (i.e., SC-FDMA sub-channels) can be encrypted using the extracted secret keys. Let $\mathbf{x}_{\text{t,un}}^{(k)}$ denote the remaining unencrypted data symbols of size $N_{\text{un}}^{(k)}$ with $N_{\text{un}}^{(k)} + N_{\text{en}}^{(k)} = N$. The $k$-th Alice then multiplexes the encrypted data symbols, $\mathbf{x}_{\text{t,en}}^{(k)}$, with $\mathbf{x}_{\text{t,un}}^{(k)}$, in the time domain to form the composite data signal, $\mathbf{x}_{\text{t,c}}^{(k)}$. Temporal AN symbols, $\mathbf{z}_{\text{t}}^{(k)}$, are then added to the entire time-domain SC-FDMA data block to secure the unencrypted data symbols. The proposed modified SC-FDMA transmitter employing the TAN-SK scheme is illustrated in Figure 3.

## A. Single-User ML Detection with Secret Keys

Using linear detectors to detect the data symbols at Eve can significantly degrade her performance since her received signal is perturbed by temporal AN interference and encryption from secret-keys. Therefore, we focus in this section on the worst-case eavesdropping scenario where Eve applies the single-user ML and multi-user ML detectors. We assume that both Bob and Eve know the indices of the encrypted and the unencrypted symbols in every SC-FDMA block. Since the encrypted symbols are perfectly secure due to the OTP encryption with secret-keys, Eve can only select the sub-channels containing the unencrypted data symbols and jointly decode them. The time-domain received signal at Eve is given by

$$\mathbf{y}_{\mathrm{Et}}^{(k)} = \sqrt{p_{\mathrm{x}}^{(k)}} \mathbf{G}^{\mathrm{c}(k)} \mathbf{x}_{\mathrm{t,c}}^{(k)} + \mathbf{S}^{(k)\top} \mathbf{F}_N^* \sum_{j=1}^{K} \mathbf{O}^{(j)} \mathbf{z}_{\mathrm{t}}^{(j)} + \mathbf{n}_{\mathrm{E}}^{(k)} \tag{43}$$

where $\mathbf{G}^{\mathrm{c}(k)} = \mathbf{F}_N^* \mathbf{G}_{\mathrm{f}}^{(k)} \mathbf{F}_N$ is the equivalent circulant channel matrix of the $A_k$-E link, $\mathbf{x}_{\mathrm{t,c}}^{(k)} \in \mathbb{C}^{N \times 1}$ is the data symbols vector which is comprised of unencrypted data symbols $\mathbf{x}_{\mathrm{t,un}}^{(k)}$ of length $N_{\mathrm{un}}^{(k)}$, in addition to the encrypted data symbols $\mathbf{x}_{\mathrm{t,en}}^{(k)}$ of length $N_{\mathrm{en}}^{(k)}$, which is given by

$$\mathbf{x}_{\mathrm{t,c}}^{(k)} = \mathbf{A}_{\mathrm{un}}^{(k)} \mathbf{x}_{\mathrm{t,un}}^{(k)} + \mathbf{A}_{\mathrm{en}}^{(k)} \mathbf{x}_{\mathrm{t,en}}^{(k)} \tag{44}$$

where the $N \times N_{\mathrm{en}}^{(k)}$ matrix $\mathbf{A}_{\mathrm{en}}^{(k)}$ and the $N \times N_{\mathrm{un}}^{(k)}$ matrix, $\mathbf{A}_{\mathrm{un}}^{(k)}$ denote the binary matrices that map the encrypted and unencrypted data symbols, respectively, to the composite data symbols vector as follows

$$\left[\mathbf{A}_{\mathrm{un}}^{(k)}\right]_{i,j} = \begin{cases} 1, & \text{unencrypted symbol } j \text{ is mapped to data symbol } i \\ 0, & \text{Otherwise} \end{cases} \tag{45}$$

---

[1]The channel-based secret-key extraction process is beyond the scope of this paper. Secret-key extraction algorithms for fading channels are given in [10], [14], where Alice and Bob 1) exchange known control and training sequences and estimate their channels. Then, they perform error correction algorithms to reconcile errors between the measurements/generated secret keys at both Alice and Bob. 3) use one-way hash functions for privacy amplification [27] to ensure that Eve does not know the final secret key.

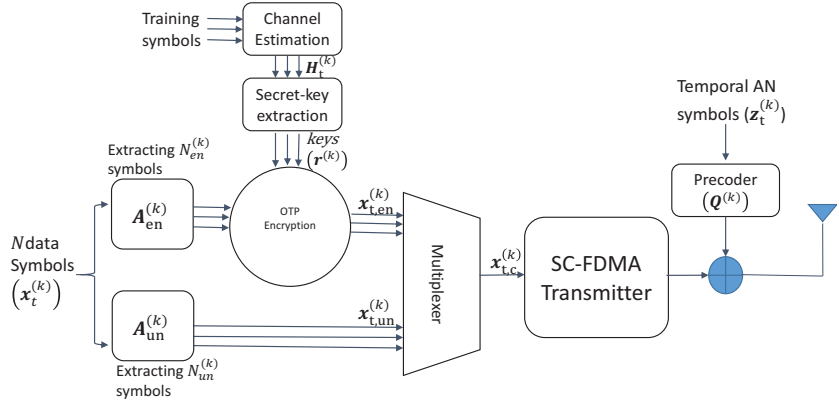

Fig. 3: $k$-th Alice's transmitter architecture to employ the hybrid TAN-SK scheme.

While extracting the unencrypted symbols, Eve's received signal is perturbed by interference from temporal AN in addition to noise due to coupling with the encrypted symbols. Therefore, the achievable rate of the Alice-Eve link using the single-user ML detection strategy is given by

$$R_{\mathrm{E}}^{(k)} = \log_2 \det \left( \mathbf{I}_N + p_{\mathrm{x}}^{(k)} \mathbf{G}^{\mathrm{c}(k)} \mathbf{A}_{\mathrm{un}}^{(k)} \mathbf{A}_{\mathrm{un}}^{(k)*} \mathbf{G}^{\mathrm{c}(k)*} \right.$$
$$\left. \times \left\{ \kappa_{\mathrm{E}} \mathbf{I}_N + p_{\mathrm{x}}^{(k)} \mathbf{G}^{\mathrm{c}(k)} \mathbf{A}_{\mathrm{en}}^{(k)} \mathbf{A}_{\mathrm{en}}^{(k)*} \mathbf{G}^{\mathrm{c}(k)*} + \mathbf{F}_N^* \mathbf{S}^{(k)\top} \left( \sum_{j=1}^{K} \mathbf{O}^{(j)} \mathbf{\Sigma}_{\mathrm{z}}^{(j)} \mathbf{O}^{(j)*} \right) \mathbf{S}^{(k)} \mathbf{F}_N \right\}^{-1} \right) \tag{46}$$

From (25), the achievable rate of the $k$-th Alice-Bob link using the MMSE detection strategy can be divided into two terms for unencrypted and encrypted symbols as follows

$$R_{\text{B}}^{(k)} = \sum_{i \in \mathcal{E}} \log_2\left(1 + \frac{p_{\text{x}}^{(k)}|\mathbf{H}_{\text{f }i,i}^{(k)}|^2}{\kappa_{\text{B}}}\right) + \sum_{i \in \mathcal{U}} \log_2\left(1 + \frac{p_{\text{x}}^{(k)}|\mathbf{H}_{\text{f }i,i}^{(k)}|^2}{\kappa_{\text{B}}}\right) \quad (47)$$

where $\mathcal{E}$ ($\mathcal{U}$) denotes the set of encrypted (unencrypted) sub-channels. The expression in (47) consists of two terms. The first term is perfectly secured due to encryption using the secret keys while the second term is unsecured and its security should be measured using the secrecy rate. Hence, the ISR of the legitimate system is given by

$$R_{\text{s}}^{(k)} = \frac{1}{M + M_{\text{cp}}}\left\{\sum_{i \in \mathcal{E}} \log_2\left(1 + \frac{p_{\text{x}}^{(k)}|\mathbf{H}_{\text{f }i,i}^{(k)}|^2}{\kappa_{\text{B}}}\right) + \left[\sum_{i \in \mathcal{U}} \log_2\left(1 + \frac{p_{\text{x}}^{(k)}|\mathbf{H}_{\text{f }i,i}^{(k)}|^2}{\kappa_{\text{B}}}\right) - R_{\text{E}}^{(k)}\right]^+\right\} \quad (48)$$

We emphasize here that in our proposed hybrid TAN-SK scheme, the term $\sum_{i \in \mathcal{E}} \log_2\left(1 + \frac{p_{\text{x}}^{(k)}|\mathbf{H}_{\text{f }i,i}^{(k)}|^2}{\kappa_{\text{B}}}\right)$ in Eqn. (48) guarantees a positive ISR in contrast to the conventional PHY security schemes. The sum ISR can be then obtained as in Eqn. (17).

### B. Multiuser ML Detection with Secret Keys

To perform multi-user ML detection over the unencrypted data symbols, Eve re-arranges the received signals as discussed in Section VI. In this case, the instantaneous sum-rate at Eve is given by

$$\sum_k R_{\text{E}}^{(k)} = \log_2 \det\left(\mathbf{I}_M + \mathbf{G}_{\text{un}}\mathbf{P}_{\text{xun}}\mathbf{G}_{\text{un}}^*\left\{\kappa_{\text{E}}\mathbf{I}_M + \mathbf{G}_{\text{en}}\mathbf{P}_{\text{xen}}\mathbf{G}_{\text{en}}^* + \left(\sum_{j=1}^{K} \mathbf{O}^{(j)}\mathbf{\Sigma}_{\text{z}}^{(j)}\mathbf{O}^{(j)*}\right)\right\}^{-1}\right) \quad (49)$$

where $\mathbf{G}_{\text{un}} \in \mathbb{C}^{M \times \left(\sum_k N_{\text{un}}^{(k)}\right)}$ and $\mathbf{G}_{\text{en}} \in \mathbb{C}^{M \times \left(\sum_k N_{\text{en}}^{(k)}\right)}$ denote the equivalent channel matrices of all-users unencrypted and encrypted data symbols, respectively, with block diagonal matrices $\mathbf{G}_{\text{f}}^{(k)}\mathbf{A}_{\text{un}}^{(k)}$ and $\mathbf{G}_{\text{f}}^{(k)}\mathbf{A}_{\text{en}}^{(k)}$, respectively. The matrices $\mathbf{P}_{\text{xun}} \in \mathbb{R}^{\left(\sum_k N_{\text{un}}^{(k)}\right) \times \left(\sum_k N_{\text{un}}^{(k)}\right)}$ and $\mathbf{P}_{\text{xen}} \in \mathbb{R}^{\left(\sum_k N_{\text{en}}^{(k)}\right) \times \left(\sum_k N_{\text{en}}^{(k)}\right)}$ denote the diagonal power matrices whose diagonal entries are the data power of all-users' unencrypted and encrypted data symbols, respectively, and are given by

$$\mathbf{P}_{\text{xun}} = \text{diag}\left(\underbrace{p_{\text{x}}^{(1)}, p_{\text{x}}^{(1)}, \ldots p_{\text{x}}^{(1)}}_{N_{\text{un}}^{(1)} \text{ terms}}, \underbrace{p_{\text{x}}^{(2)}, \ldots p_{\text{x}}^{(2)}}_{N_{\text{un}}^{(2)} \text{ terms}}, \ldots, \underbrace{p_{\text{x}}^{(K)}, \ldots p_{\text{x}}^{(K)}}_{N_{\text{un}}^{(K)} \text{ terms}}\right) \quad (50)$$

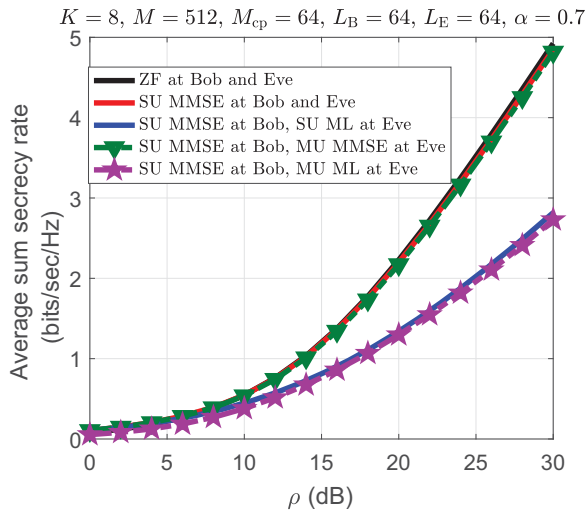

Fig. 4: Average sum secrecy rate versus SNR for different detection strategies.

$$\mathbf{P}_{\text{xen}} = \text{diag} \left( \underbrace{p_{\text{x}}^{(1)}, p_{\text{x}}^{(1)}, ... p_{\text{x}}^{(1)}}_{N_{\text{en}}^{(1)} \text{ terms}}, \underbrace{p_{\text{x}}^{(2)}, ... p_{\text{x}}^{(2)}}_{N_{\text{en}}^{(2)} \text{ terms}}, ..., \underbrace{p_{\text{x}}^{(K)}, ... p_{\text{x}}^{(K)}}_{N_{\text{en}}^{(K)} \text{ terms}} \right) \tag{51}$$

The sum ISR can be given by

$$R_{\text{s}} = \frac{1}{M + M_{\text{cp}}} \left\{ \sum_k \sum_{i \in \mathcal{E}} \log_2 \left( 1 + \frac{p_{\text{x}}^{(k)} |\mathbf{H}_{\text{f}\,i,i}^{(k)}|^2}{\kappa_{\text{B}}} \right) + \left[ \sum_k \sum_{i \in \mathcal{U}} \log_2 \left( 1 + \frac{p_{\text{x}}^{(k)} |\mathbf{H}_{\text{f}\,i,i}^{(k)}|^2}{\kappa_{\text{B}}} \right) - \sum_k R_{\text{E}}^{(k)} \right]^+ \right\} \tag{52}$$

**Remark 2.** *The Alices-Eve's rate expressions in (46) and (49) are based on the assumption that Eve knows the AN precoding matrices $\mathbf{Q}^{(k)}$ $\forall k$. Given that the null space of a matrix can be obtained from the singular value decomposition (SVD) of that matrix (i.e., by selecting the right singular vectors corresponding to zero singular values), knowledge of the AN precoding matrix, $\mathbf{Q}^{(k)}$, may imply that Eve knows some information about the Alices-Bob channel matrices. This, in turn, might sacrifice the key's security which is based on the CIR of the Alices-Bob links. However, as shown in Lemma 1 of [26], the right singular vectors of a random matrix do not reveal any information about the matrix itself. Hence, even when Eve knows the AN precoding matrix, she does not know the matrix itself and this information does not reveal additional correlated information about the CSI of the Alices-Bob links. We can therefore assume that Eve has full knowledge of the null space matrices without compromising the information-theoretic security.*

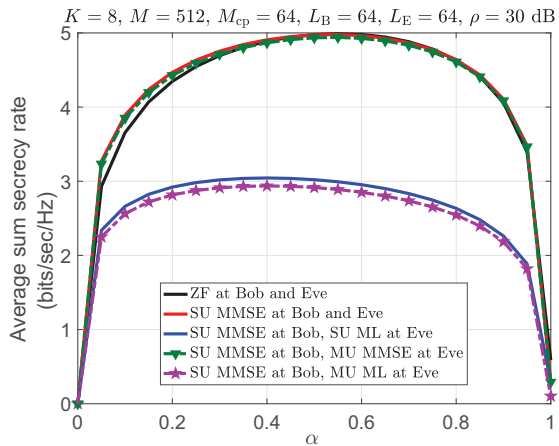

Fig. 5: Average sum secrecy rate versus the power fraction allocation parameter, $\alpha$.

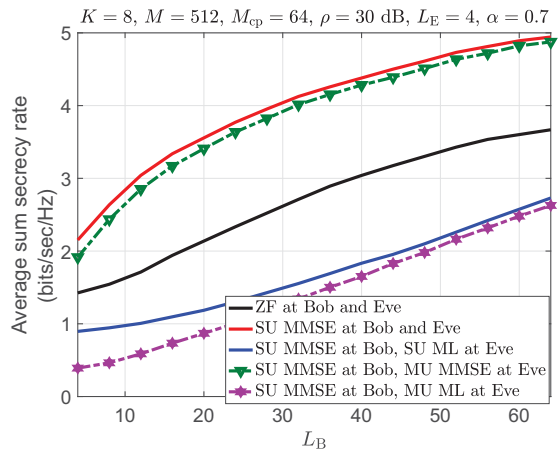

Fig. 6: Average sum secrecy rate versus Alices-Bob's channel memory, $L_\text{B}$.

## VIII. SIMULATION RESULTS

In this section, we evaluate the average secrecy rate performance of the temporal-AN scheme in SC-FDMA systems under the investigated detection strategies. Unless stated explicitly, we consider an SC-FDMA system with $K = 8$ users, $M = 512$, $N = 64$, $M_\text{cp} = 64$, $L_\text{B} = L_\text{B}^{(k)} = 64$ $\forall\, k$, $L_\text{E} = L_\text{E}^{(k)} = 64$ $\forall\, k$, and $\rho = \frac{p_\text{t}}{N\kappa_\text{B}} = \frac{p_\text{t}}{N\kappa_\text{E}} = 30$ dB. Each CIR tap has an average power of $\sigma_{\text{A}-\text{B}}^{2(k)} = 1/(L_\text{B}^{(k)} + 1)$ and $\sigma_{\text{A}-\text{E}}^{2(k)} = 1/(L_\text{E}^{(k)} + 1)$ $\forall k$, (i.e., uniform power delay profile). We assume $\alpha = \alpha^{(k)} = 0.7$ $\forall\, k$, which means that $70\%$ of the total transmit power is assigned to data signals and the remaining $30\%$ is assigned to AN signals. The data power is divided equally across the data symbols and the AN power is divided equally across the AN symbols.

Fig. 4 depicts the achieved average sum secrecy rate using the temporal-AN scheme versus the input SNR when Eve follows different detection strategies (single-user (SU) ZF, SU MMSE, multiuser (MU) MMSE, SU ML, and MU ML) while Bob is constrained to simple linear detection strategies (SU ZF and SU MMSE). As Bob and Eve increase their detector complexity from ZF to MMSE, both of them achieve gains in their link rates. That is why SU ZF and SU MMSE achieve close secrecy rate performance. As Eve adopts the ML detection strategy, she exploits the correlation properties of the temporal AN to minimize its effect as explained in Remark 1. As a result, Eve achieves gains in her link rate and the average secrecy rate decreases. Fig. 4 also reveals that the average secrecy rate gains Eve achieves by adopting MU detection strategies (MMSE or ML) result in a slight degradation in the achieved secrecy rate. This demonstrates the robustness of the proposed temporal AN scheme to high-complexity detectors at Eve. Even when Eve has unlimited computational capabilities while Bob maintains simple per-sub-channel detection schemes (SU ZF and SU MMSE), a high average secrecy rate is achieved.

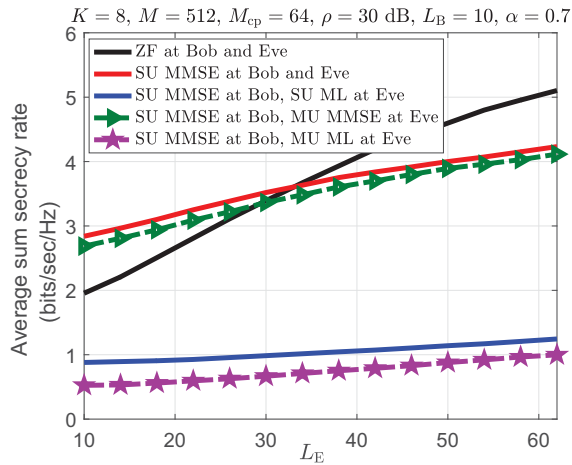

Fig. 7: Average sum secrecy rate versus Alices-Eve's channel memory, $L_\text{E}$.

In Fig. 5, we show the achieved average secrecy rate versus the transmit power fraction parameter, $\alpha$, for different detection strategies at Bob and Eve. The figure demonstrates that when Eve adopts the ML detector, the average secrecy rate does not vary as long as $\alpha$ is not close to zero or one which corroborates our analytical results in Proposition 2. The case of $\alpha = 1$ corresponds to the benchmark case of no-AN injection. Therefore, Fig. 5 quantifies the average secrecy rate gain of our proposed temporal-AN scheme compared to the no-AN injection scenario.

Figs 6 and 7 depict the achieved average secrecy rates for different detection strategies as $L_\text{B}$ and $L_\text{E}$ increase, respectively. As discussed in Proposition 1, the number of useful AN streams depends on the channel memories of the Alices-Bob and Alices-Eve links. Dispersive channels can accommodate more useful AN streams. Increasing the number of useful AN streams introduces diversity in interference and reduces the AN correlation at Eve. Therefore, the Alices-Eve link rates experience more degradation and the secrecy rate is boosted.

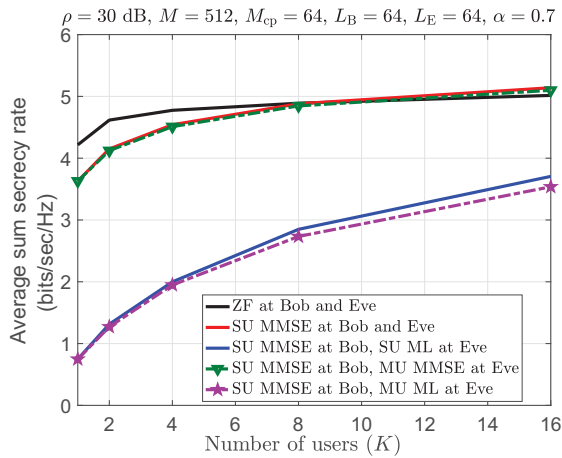

Fig. 8: Average sum secrecy rate versus the number of users, $K$.

Fig 8 depicts the average secrecy rate performance for different detection strategies versus the number of users, $K$. We fix $M$ and $M_{\text{cp}}$ while $N = M/K$ varies as we vary $K$. Each user transmits a temporal AN signal that not only interferes with Eve's reception of the user's own data signal, but it interferes with Eve's reception of the other users' signals as well. This, in turn, increases the achieved average secrecy rate even when Eve adopts the MU ML detection strategy. This confirms the robustness of our proposed temporal-AN scheme in highly-dense user deployments.

Fig 9 depicts the average sum secrecy rate performance of the hybrid TAN-SK scheme versus the number of encrypted symbols, $N_{\text{en}} = N_{\text{en}}^{(k)} \, \forall k$, for the cases of SU and MU ML detectors at Eve. As the number of encrypted symbols increases, the gain of the perfectly secured term in (48) and (52) increases and the secrecy rate is boosted. To highlight the secrecy rate gains, the performance is compared with the temporal-AN-only scheme and the benchmark case with no secret-keys generation or temporal AN. In Fig 10, we show the average sum secrecy rate of the hybrid TAN-SK scheme vs $\alpha$. The figure quantifies the performance gains of the hybrid TAN-SK scheme and illustrates the gains introduced by the temporal-AN only and secret-keys only schemes for $N_{\text{en}} = 10$ encrypted data symbols. The performance is compared with the benchmark case of no injected temporal-AN or secret-keys generation.

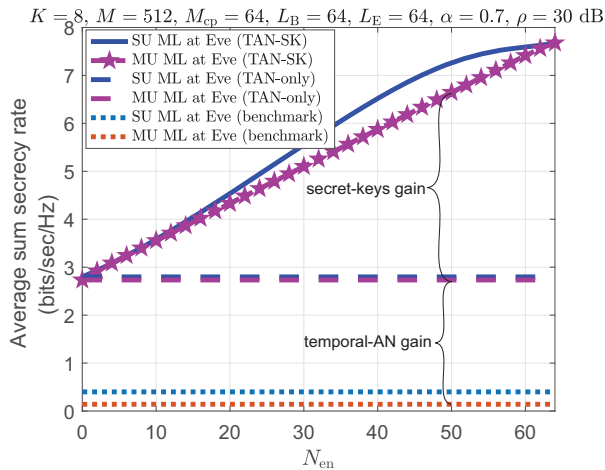

Fig. 9: Average sum secrecy rate of the hybrid TAN-SK scheme vs the number of encrypted symbols, $N_{\text{en}}$.

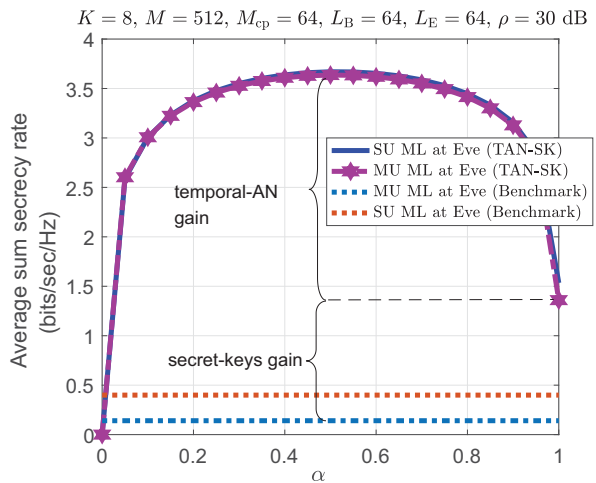

Fig. 10: Average sum secrecy rate of the hybrid TAN-SK scheme vs the power fraction allocation parameter, $\alpha$.

## IX. CONCLUSIONS

We proposed a temporal-AN-aided scheme to secure SC-FDMA communications from eavesdropping. We analyzed the achieved data rates of the Alices-Bob links (legitimate links) and Alices-Eve links (eavesdropping links) under low-complexity detectors at Bob and high-complexity detectors with global channel knowledge at Eve. In addition, we showed that, as Eve increases her SU detector complexity (from ZF to MMSE and to ML) to exploit the correlation properties of the temporal AN signal, she can reduce the AN effect on her data rate. As Eve adopts the high-complexity multiuser detection strategies, she can only introduce slight degradation in the achieved secrecy rate which validates the robustness of our proposed scheme. In addition, we investigated the performance of the temporal-AN scheme under different transmit power allocation fractions for the data and AN signals. Furthermore, we proved that the number of useful AN streams that can be injected to degrade Eve's SNR is equal to the maximum of the $k$-th Alice-Bob and the $k$-th Alice-Eve channel memories. We derived a closed-form expression for a lower bound on the average secrecy rate at high input SNR levels. The derived expression showed that, in the worst-case secrecy scenario for the legitimate system when Eve uses the ML detector, the average secrecy rate is always positive and increases linearly with the number of useful AN streams. We showed that increasing the number of users increases the achievable secrecy rate due to the ability of those users to transmit AN signals that degrade Eve's receiver only. Moreover, for the case of partial Alices-Bob channel knowledge at Eve, we proposed a hybrid TAN-SK scheme to exploit Eve's unawareness of the Alices-Bob channel in enhancing the secrecy rate. Our proposed TAN-SK scheme guarantees a positive instantaneous secrecy rate even when Eve performs the high-complexity (SU or MU) ML detector. Simulations results demonstrated that for the case of global channel knowledge at Eve, adopting the proposed temporal-AN injection scheme is critical to achieve higher secrecy rates compared to the benchmark case with no-AN injection. For the case of partial Alices-Bob channels knowledge at Eve, additional secrecy rate gains are achievable through our proposed TAN-SK scheme.

## APPENDIX A
## PROOF OF PROPOSITION 1

The $k$-th Alice superimposes the temporal-AN signal on the data signal after CP insertion. The composite (data+AN) signal is received at Bob after it passes through the time-domain $\text{A}_k - \text{B}$ channel matrix given by

$$\mathbf{R}^{\text{cp}}\mathbf{H}_{\text{t}}^{(k)} = \left[\mathbf{0}_{M\times(M_{\text{cp}}-L_{\text{B}}^{(k)})} \ \mathbf{H}'^{(k)}_{M\times(M+L_{\text{B}}^{(k)})}\right] \tag{53}$$

where $\mathbf{H}'^{(k)} \in \mathbb{C}^{M\times(M+L_{\text{B}}^{(k)})}$ is the Toeplitz upper-triangular $\text{A}_k - \text{B}$ channel matrix after CP removal with $\left[h'^{(k)}(0), h'^{(k)}(1), \ldots, h'^{(k)}(L_{\text{B}}^{(k)}), 0, \ldots, 0\right]$ as its first row.

The AN signal is precoded using a precoding matrix $\mathbf{Q}^{(k)}$ which is designed to ensure that the temporal AN is eliminated at Bob by spanning the null-space of $\mathbf{R}^{\text{cp}}\mathbf{H}_{\text{t}}^{(k)}$ as follows

$$\begin{aligned}\mathbf{Q}^{(k)} &= \text{Null}\left(\mathbf{R}^{\text{cp}}\mathbf{H}^{\text{time}(k)}\right) \\ &= \text{Null}\left(\left[\mathbf{0}_{M\times\left(M_{\text{cp}}-L_{\text{B}}^{(k)}\right)} \ \mathbf{H}'^{(k)}_{M\times\left(M+L_{\text{B}}^{(k)}\right)}\right]\right)\end{aligned} \tag{54}$$

We notice that $\mathbf{R}^{\text{cp}}\mathbf{H}^{\text{time}(k)}$ has $\left(M_{\text{cp}} - L_{\text{B}}^{(k)}\right)$ all-zero columns. Let $\mathbf{W} = \left[\mathbf{I}_{\left(M_{\text{cp}}-L_{\text{B}}^{(k)}\right)} \ \mathbf{0}_{\left(M+L_{\text{B}}^{(k)}\right)\times\left(M_{\text{cp}}-L_{\text{B}}^{(k)}\right)}\right]^{\top}$ denote the orthonormal basis matrix that extracts the all-zero columns out of $\mathbf{R}^{\text{cp}}\mathbf{H}^{\text{time}(k)}$ as follows

$$\mathbf{R}^{\text{cp}}\mathbf{H}^{\text{time}(k)}\mathbf{W} = \mathbf{0}_{M\times\left(M_{\text{cp}}-L_{\text{B}}^{(k)}\right)} \tag{55}$$

The null space precoder, $\mathbf{Q}^{(k)}$, can therefore be expressed as

$$\mathbf{Q}^{(k)} = \left[\mathbf{W}_{(M+M_{\text{cp}})\times(M_{\text{cp}}-L_{\text{B}})} \ \mathbf{Q}^{'(k)}_{(M+M_{\text{cp}})\times L_{\text{B}}^{(k)}}\right] \tag{56}$$

where $\mathbf{Q}^{'(k)} = \text{Null}\left(\mathbf{H}^{'(k)}\right)$ denotes the orthonormal basis matrix that spans the null space of $\mathbf{H}^{'(k)}$. At Eve, the $k$-th Alice AN signal vector interfering with the $l$-th Alice $N$ data sub-channels is given by

$$\mathbf{i}_{N\times 1}^{(k)} = \mathbf{S}^{(l)\top}\mathbf{F}_M\mathbf{R}^{\text{cp}}\mathbf{G}_{\text{t}}^{(j)}\mathbf{Q}^{(j)}\mathbf{z}_{\text{t}}^{(k)} \tag{57}$$

The equivalent $A_k - E$ time-domain channel matrix after CP removal is similarly given by

$$\mathbf{R}^{\text{cp}}\mathbf{G}^{\text{time}(k)} = \left[\mathbf{0}_{M\times(M_{\text{cp}}-L_{\text{E}}^{(k)})} \ \mathbf{G}^{'(k)}_{M\times(M+L_{\text{E}}^{(k)})}\right] \tag{58}$$

where $\mathbf{G}^{'(k)} \in \mathbb{C}^{M\times(M+L_{\text{E}}^{(k)})}$ is the Toeplitz upper-triangular $A_k - E$ channel matrix. The AN vector across the data SC-FDMA sub-channels of Eve can therefore be expressed as

$$\begin{aligned}\mathbf{i}_{N\times 1}^{(k)} &= \mathbf{S}^{(l)\top}\mathbf{F}_M\left[\mathbf{0}_{M\times(M_{\text{cp}}-L_{\text{E}}^{(k)})} \ \mathbf{G}^{'(k)}_{M\times(M+L_{\text{E}}^{(k)})}\right]\times\left[\mathbf{W}_{(M+M_{\text{cp}})\times(M_{\text{cp}}-L_{\text{B}}^{(k)})} \ \mathbf{Q}^{'(k)}_{(M+M_{\text{cp}})\times L_{\text{B}}^{(k)}}\right] \mathbf{z}_{\text{t}\,M_{\text{cp}}\times 1}^{(k)}\\ &= \left[\mathbf{0}_{N\times\left(M_{\text{cp}}-L_{\text{u}}^{(k)}\right)} \ \mathbf{U}^{'(k)}_{\text{useful}\,\left(N\times L_{\text{u}}^{(k)}\right)}\right] \times \mathbf{z}_{\text{t}\,M_{\text{cp}}\times 1}^{(k)}\end{aligned} \tag{59}$$

where $L_{\text{u}}^{(k)} = \max\left(L_{\text{B}}^{(k)}, L_{\text{E}}^{(k)}\right)$ and $\mathbf{U}^{'(k)}_{\text{useful}}$ is given by,

$$\mathbf{U}^{'(k)}_{\text{useful}} = \mathbf{S}^{(l)\top}\mathbf{F}_M\left[\mathbf{0}_{\left(M\times(M_{\text{cp}}-L_{\text{E}}^{(k)})\right)} \ \mathbf{G}^{'(k)}_{M\times(M+L_{\text{E}}^{(k)})}\right]\times\left[\mathbf{W}_{(M+M_{\text{cp}})\times\left(L_{\text{u}}^{(k)}-L_{\text{B}}^{(k)}\right)} \ \mathbf{Q}^{'(k)}_{(M+M_{\text{cp}})\times L_{\text{B}}^{(k)}}\right].$$

From Eqn. (59), we notice that the first $\left(M_{\text{cp}} - L_{\text{u}}^{(k)}\right)$ AN streams of $\mathbf{z}_{\text{t}\,M_{\text{cp}}\times 1}^{(k)}$ are projected into all-zero columns. That is, those AN streams lie in the $A_k - E$ channel null space and cause no harm to Eve. The matrix $\left(\mathbf{R}^{\text{cp}}\mathbf{G}_{\text{t}}^{(k)}\mathbf{Q}^{(k)}\right)$ has a rank of $L_{\text{u}}^{(k)}$. Therefore, there are only $L_{\text{u}}^{(k)}$ *useful* AN directions in the sense that they can degrade Eve's SINR.

We conclude that the number of useful AN streams depend on the maximum of the $A_k - B$ and $A_k - E$ channels memories. As $L_{\text{B}}^{(k)}$ increases, the number of orthonormal basis vectors of the $A_k - B$ channel null space increases which, in turn, increases the number of AN streams. As $L_{\text{E}}^{(k)}$ increases, the $A_k - E$ channel spreads each AN stream across more of its data sub-channels.

# APPENDIX B
## PROOF OF PROPOSITION 2

Assuming ML detection, the instantaneous data rate expression of the $A_k-E$ link is given by

$$R_{\rm E}^{(k)} = \log_2 \det \left( \mathbf{I}_N + p_{\rm x}^{(k)} \mathbf{G}_{\rm f}^{(k)} \mathbf{G}_{\rm f}^{(k)*} \left( \kappa_{\rm E} \mathbf{I}_N + \left( \sum_{j=1}^{K} \mathbf{S}^{(k)\top} \mathbf{O}^{(j)} \boldsymbol{\Sigma}_{\rm z}^{(j)} \mathbf{O}^{(j)*} \mathbf{S}^{(k)} \right) \right)^{-1} \right) \quad (60)$$

By dividing the data power equally across the $N$ data symbols, each symbol will have a power of $p_{\rm x}^{(k)} = \frac{\alpha^{(k)} p_{\rm t}}{N}$. Similarly, by performing equal power allocation across the $L_{\rm u}^{(k)}$ AN symbols, each symbol will have a power of $p_{\rm z}^{(j)} = (1 - \alpha^{(j)}) \frac{p_{\rm t}}{L_{\rm u}^{(j)}}$.

$$R_{\rm E}^{(k)} = \log_2 \det \left( \mathbf{I}_N + \alpha^{(k)} \frac{p_{\rm t}}{N} \mathbf{G}_{\rm f}^{(k)} \mathbf{G}_{\rm f}^{(k)*} \left( \kappa_{\rm E} \mathbf{I}_N + \left( \sum_{j=1}^{K} (1-\alpha^{(j)}) \frac{p_{\rm t}}{L_{\rm u}^{(j)}} \mathbf{S}^{(k)\top} \mathbf{O}^{(j)} \mathbf{O}^{(j)*} \mathbf{S}^{(k)} \right) \right)^{-1} \right) \quad (61)$$

Since the matrix $\left( \mathbf{S}^{(k)\top} \mathbf{O}^{(j)} \mathbf{O}^{(j)*} \mathbf{S}^{(k)} \right)$ is positive semi-definite, the term $\left( \sum_{j=1}^{K} p_{\rm z}^{(j)} \mathbf{S}^{(k)\top} \mathbf{O}^{(j)} \mathbf{O}^{(j)*} \mathbf{S}^{(k)} \right)$ can be factored as $\left( \sum_{j=1}^{K} p_{\rm z}^{(j)} \mathbf{S}^{(k)\top} \mathbf{O}^{(j)} \mathbf{O}^{(j)*} \mathbf{S}^{(k)} \right) = \tilde{\mathbf{Q}} \tilde{\mathbf{Q}}^*$ where $\tilde{\mathbf{Q}}$ in an $N \times N$ matrix which represents the combined AN term at Eve. By performing the SVD on $\tilde{\mathbf{Q}}$, we get, $\tilde{\mathbf{Q}} = \mathbf{U}_{\rm z} \boldsymbol{\Lambda}_{\rm z} \mathbf{V}_{\rm z}^*$, where $\boldsymbol{\Lambda}_{\rm z}$ is an $N \times N$ diagonal matrix with at least $L_{\rm u}^{\max} = \max_k \{L_{\rm u}^{(k)}\}$ non-zero singular values. The instantaneous data rate of the $A_k-E$ link is given by

$$\begin{aligned} R_{\rm E}^{(k)} &= \log_2 \det \left( \mathbf{I}_N + p_{\rm x}^{(k)} \mathbf{G}_{\rm f}^{(k)} \mathbf{G}_{\rm f}^{(k)*} \left( \kappa_{\rm E} \mathbf{I}_N + \mathbf{U}_{\rm z} \boldsymbol{\Lambda}_{\rm z} \boldsymbol{\Lambda}_{\rm z}^* \mathbf{U}_{\rm z}^* \right)^{-1} \right) \\ &= \log_2 \det \left( \mathbf{I}_N + p_{\rm x}^{(k)} \mathbf{G}_{\rm f}^{(k)} \mathbf{G}_{\rm f}^{(k)*} \mathbf{U}_{\rm z} \left( \kappa_{\rm E} \mathbf{I}_N + \boldsymbol{\Lambda}_{\rm z} \boldsymbol{\Lambda}_{\rm z}^* \right)^{-1} \mathbf{U}_{\rm z}^* \right) \quad (62) \\ &= \log_2 \det \left( \mathbf{I}_N + p_{\rm x}^{(k)} \mathbf{U}_{\rm z}^* \mathbf{G}_{\rm f}^{(k)} \mathbf{G}_{\rm f}^{(k)*} \mathbf{U}_{\rm z} \left( \kappa_{\rm E} \mathbf{I}_N + \boldsymbol{\Lambda}_{\rm z} \boldsymbol{\Lambda}_{\rm z}^* \right)^{-1} \right) \end{aligned}$$

The last equality holds due to Sylvester's identity. Notice that the diagonal matrix $\boldsymbol{\Lambda}_{\rm z} \boldsymbol{\Lambda}_{\rm z}^*$ represents the AN power at Eve. Define the AN-plus-AWGN covariance matrix, $\mathbf{V} = (\kappa_{\rm E} \mathbf{I}_N + \boldsymbol{\Lambda}_{\rm z} \boldsymbol{\Lambda}_{\rm z}^*)$, then the $k$-th Alice-Eve's instantaneous data rate is given by

$$\begin{aligned} R_{\rm E}^{(k)} &= \log_2 \det \left( \mathbf{I}_N + p_{\rm x}^{(k)} \mathbf{U}_{\rm z}^* \mathbf{G}_{\rm f}^{(k)} \mathbf{G}_{\rm f}^{(k)*} \mathbf{U}_{\rm z} \mathbf{V}^{-1} \right) \\ &= \log_2 \det \left( \mathbf{V} + p_{\rm x}^{(k)} \mathbf{U}_{\rm z}^* \mathbf{G}_{\rm f}^{(k)} \mathbf{G}_{\rm f}^{(k)*} \mathbf{U}_{\rm z} \right) - \log_2 \det (\mathbf{V}) \end{aligned} \quad (63)$$

The matrix $\mathbf{V} + p_{\rm x}^{(k)} \mathbf{U}_{\rm z}^* \mathbf{G}_{\rm f}^{(k)} \mathbf{G}_{\rm f}^{(k)*} \mathbf{U}_{\rm z}$ is Hermitian, hence, we can apply the Hadamard inequality to derive an upper bound on the data rate of the Alice-Eve link as follows

$$R_{\rm E}^{(k)} \leq \sum_{i=1}^{N} \log_2 \left( [\mathbf{V}]_{i,i} + p_{\rm x}^{(k)} [\mathbf{U}_{\rm z}^* \mathbf{G}_{\rm f}^{(k)} \mathbf{G}_{\rm f}^{(k)*} \mathbf{U}_{\rm z}]_{i,i} \right) - \log_2 \det (\mathbf{V}) \quad (64)$$

Defining the $i$-th row of $\mathbf{U}_{\rm z}^*$ as $\mathbf{u}_i = [u_{i,1}, u_{i,2}, \cdots, u_{i,N}]$, the $i$-th diagonal element of $\mathbf{U}_{\rm z}^* \mathbf{G}_{\rm f}^{(k)} \mathbf{G}_{\rm f}^{(k)*} \mathbf{U}_{\rm z}$ is thus given by $[\mathbf{U}_{\rm z}^* \mathbf{G}_{\rm f}^{(k)} \mathbf{G}_{\rm f}^{(k)*} \mathbf{U}_{\rm z}]_{i,i} = \sum_{j=1}^{N} |u_{i,j}|^2 |G_j^{(k)}|^2$ with $G_j^{(k)}$ denoting the $j$-th frequency-

domain channel coefficient of the $k$-th Alice-Eve link. Applying Jensen's inequality to the concave function $\sum_{i=1}^{N} \log_2 \left( [\mathbf{V}]_{i,i} + p_{\mathrm{x}}^{(k)} [\mathbf{U}_{\mathrm{z}}^* \mathbf{G}_{\mathrm{f}}^{(k)} \mathbf{G}_{\mathrm{f}}^{(k)*} \mathbf{U}_{\mathrm{z}}]_{i,i} \right)$, the achieved average rate of the $k$-th Alice-Eve link can be upper-bounded as follows

$$\mathbb{E}\{R_{\mathrm{E}}^{(k)}\} \leq \sum_{i=1}^{N} \mathbb{E} \left\{ \log_2 \left( [\mathbf{V}]_{i,i} + p_{\mathrm{x}}^{(k)} \mathbb{E}_{\mathrm{G}^{(\mathrm{k})}} \left\{ [\mathbf{U}_{\mathrm{z}}^* \mathbf{G}_{\mathrm{f}}^{(k)} \mathbf{G}_{\mathrm{f}}^{(k)*} \mathbf{U}_{\mathrm{z}}]_{i,i} \right\} \right) \right\} - \mathbb{E}\{\log_2 \det(\mathbf{V})\}$$

$$= \sum_{i=1}^{N} \mathbb{E} \left\{ \log_2 \det \left( [\mathbf{V}]_{i,i} + p_{\mathrm{x}}^{(k)} \times \mathbb{E}_{\mathrm{G}^{(\mathrm{k})}} \left\{ \sum_{j=1}^{N} |u_{i,j}|^2 |G_j^{(k)}|^2 \right\} \right) \right\} - \mathbb{E}\{\log_2 \det(\mathbf{V})\} \quad (65)$$

where we averaged only over the channel $G_j^{(k)}$ and kept the expectation over the other random variables to obtain a tight bound. That is, the outside expectation is over all random variables except $G_j^{(k)}$. Due to the channel independence from one link to another, the random variable $|G_j^{(k)}|^2$ is independent from $|u_{i,1}|^2$. Hence, we have

$$\mathbb{E}_{\mathrm{G}_{j}^{(\mathrm{k})}} \left\{ \sum_{j=1}^{N} |u_{i,j}|^2 |\mathrm{G}_j^{(\mathrm{k})}|^2 \right\} = \left\{ \sum_{j=1}^{N} |u_{i,j}|^2 \mathbb{E}_{\mathrm{G}^{(\mathrm{k})}} \{|G_j^{(k)}|^2\} \right\} = \left\{ \sum_{j=1}^{N} |u_{i,j}|^2 \sigma_{\mathrm{A-E}}^{(k)\,2} (L_{\mathrm{E}}^{(k)} + 1) \right\} \quad (66)$$

where $\mathbb{E}_{\mathrm{G}_{j}^{(\mathrm{k})}}\{|G_j^{(k)}|^2\} = \mathbb{E}\{\sum_{i=1}^{L_{\mathrm{E}}^{(k)}} |g(i)|^2\} = \sigma_{\mathrm{A-E}}^{(k)\,2}(L_{\mathrm{E}}^{(k)} + 1)$ with $g(i)$ denoting the $i$-th tap of the Alice-Eve link CIR. Accordingly,

$$\mathbb{E}\{R_{\mathrm{E}}^{(k)}\} \leq \sum_{i=1}^{N} \mathbb{E} \left\{ \log_2 \left( [\mathbf{V}]_{i,i} + p_{\mathrm{x}}^{(k)} \sigma_{\mathrm{A-E}}^{(k)\,2}(L_{\mathrm{E}}^{(k)} + 1) \times \sum_{j=1}^{N} |u_{i,j}|^2 \right) \right\} - \mathbb{E}\{\log_2 \det(\mathbf{V})\} \quad (67)$$

Let $\sigma_i^2$ denote the $i$-th diagonal entry of the diagonal matrix $\mathbf{V}$. Considering the worst-case scenario where $\tilde{\mathbf{Q}}$ has only $L_{\mathrm{u}}^{\max}$ non-zero singular values, $\sigma_i^2$, is given by

$$\sigma_i^2 = [\mathbf{V}]_{i,i} = \begin{cases} \kappa_{\mathrm{E}} + \delta_{z,i}^2, & i \leq L_{\mathrm{u}}^{\max} \\ \kappa_{\mathrm{E}}, & \text{otherwise} \end{cases} \quad (68)$$

where $\delta_{z,i}$ denotes the $i$-th diagonal entry of $\mathbf{\Lambda}_{\mathrm{z}}$ and $\delta_{z,i}^2$ represents the AN power at Eve which can be expressed as a fraction of the total transmit power. Hence, $\delta_{z,i}^2 = \beta_i K p_t$ where $0 \leq \beta_i \leq 1, \forall i$. On the other hand, $\det(\mathbf{V}) = \prod_{i=1}^{N} [\mathbf{V}]_{i,i} = \prod_{i=1}^{N} \sigma_i^2$. Since $\mathbf{U}_z$ is unitary, $\sum_{j=1}^{N} |u_{i,j}|^2 = \sum_{i=1}^{N} |u_{i,j}|^2 = 1$. Hence,

$$\mathbb{E}\{R_{\mathrm{E}}^{(k)}\} \leq \sum_{i=1}^{N} \mathbb{E} \left\{ \log_2 \left( 1 + \frac{p_{\mathrm{x}}^{(k)} \sigma_{\mathrm{A-E}}^{(k)\,2}(L_{\mathrm{E}}^{(k)} + 1)}{\sigma_i^2} \right) \right\} \quad (69)$$

The average secrecy rate is thus lower bounded by

$$\mathbb{E}\{R_{\mathrm{B}}^{(k)}\} - \mathbb{E}\{R_{\mathrm{E}}^{(k)}\} \geq \sum_{i=1}^{N} \mathbb{E} \left\{ \log_2 \left( 1 + \frac{|\mathbf{H}_{\mathrm{f}\,i,i}^{(k)}|^2 p_{\mathrm{x}}^{(k)}}{\kappa_{\mathrm{B}}} \right) \right\} - \sum_{i=1}^{N} \mathbb{E} \left\{ \log_2 \left( 1 + \frac{p_{\mathrm{x}}^{(k)} \sigma_{\mathrm{A-E}}^{(k)\,2}(L_{\mathrm{E}}^{(k)} + 1)}{\sigma_i^2} \right) \right\} \quad (70)$$

Splitting each summation into two sums from 1 to $L_{\mathrm{u}}^{\max}$ and from $L_{\mathrm{u}}^{\max}+1$ to $N$, we get

$$\mathbb{E}\{R_{\mathrm{B}}^{(k)}\} - \mathbb{E}\{R_{\mathrm{E}}^{(k)}\} \geq \mathbb{E}\left\{\sum_{i=1}^{L_{\mathrm{u}}^{\max}} \log_2\left(1 + \frac{|\mathbf{H}_{\mathrm{f}\,i,i}^{(k)}|^2 p_{\mathrm{x}}^{(k)}}{\kappa_{\mathrm{B}}}\right) + \sum_{i=L_{\mathrm{u}}^{\max}+1}^{N} \log_2\left(1 + \frac{|\mathbf{H}_{\mathrm{f}\,i,i}^{(k)}|^2 p_{\mathrm{x}}^{(k)}}{\kappa_{\mathrm{B}}}\right)\right.$$
$$\left. - \sum_{i=1}^{L_{\mathrm{u}}^{\max}} \log_2\left(1 + \frac{p_{\mathrm{x}}^{(k)} \sigma_{\mathrm{A-E}}^{(k)\,2}(L_{\mathrm{E}}^{(k)}+1)}{\sigma_i^2}\right) - \sum_{i=L_{\mathrm{u}}^{\max}+1}^{N} \log_2\left(1 + \frac{p_{\mathrm{x}}^{(k)} \sigma_{\mathrm{A-E}}^{(k)\,2}(L_{\mathrm{E}}^{(k)}+1)}{\sigma_i^2}\right)\right\} \tag{71}$$

At very high input SNR levels $\sigma_i^2 \approx \delta_{z,i}^2$ for $i \leq L_{\mathrm{u}}^{\max}$ and the average secrecy rate is given by

$$\mathbb{E}\{R_{\mathrm{B}}^{(k)}\} - \mathbb{E}\{R_{\mathrm{E}}^{(k)}\} \geq \mathbb{E}\left\{\sum_{i=1}^{L_{\mathrm{u}}^{\max}} \log_2\left(\frac{\alpha^{(k)} p_t |\mathbf{H}_{\mathrm{f}\,i,i}^{(k)}|^2}{N \kappa_{\mathrm{B}}}\right) + \sum_{i=L_{\mathrm{u}}^{\max}+1}^{N} \log_2\left(\frac{\alpha^{(k)} p_t |\mathbf{H}_{\mathrm{f}\,i,i}^{(k)}|^2}{N \kappa_{\mathrm{B}}}\right)\right.$$
$$\left. - \sum_{i=1}^{L_{\mathrm{u}}^{\max}} \log_2\left(1 + \frac{\alpha^{(k)} \frac{p_t}{N} \sigma_{\mathrm{A-E}}^2 (L_{\mathrm{E}}+1)}{\beta_i\, K\, p_t}\right) - \sum_{i=L_{\mathrm{u}}^{\max}+1}^{N} \log_2\left(\frac{\alpha^{(k)} p_t \sigma_{\mathrm{A-E}}^2 (L_{\mathrm{E}}+1)}{N \kappa_{\mathrm{E}}}\right)\right\} \tag{72}$$

Using the logarithmic function properties, we get

$$\mathbb{E}\{R_{\mathrm{B}}^{(k)}\} - \mathbb{E}\{R_{\mathrm{E}}^{(k)}\} \geq \mathbb{E}\left\{\sum_{i=1}^{L_{\mathrm{u}}^{\max}} \log_2\left(\frac{\alpha^{(k)} p_t |\mathbf{H}_{\mathrm{f}\,i,i}^{(k)}|^2}{N \kappa_{\mathrm{B}}}\right) + \sum_{i=L_{\mathrm{u}}^{\max}+1}^{N} \log_2\left(\frac{\frac{|\mathbf{H}_{\mathrm{f}\,i,i}^{(k)}|^2}{\kappa_{\mathrm{B}}}}{\frac{\sigma_{\mathrm{A-E}}^{(k)\,2}(L_{\mathrm{E}}^{(k)}+1)}{\kappa_{\mathrm{E}}}}\right)\right.$$
$$\left. - \sum_{i=1}^{L_{\mathrm{u}}^{\max}} \log_2\left(1 + \frac{\frac{\alpha^{(k)}}{N} \sigma_{\mathrm{A-E}}^{(k)\,2}(L_{\mathrm{E}}^{(k)}+1)}{\beta_i\, K}\right)\right\} \tag{73}$$

This can be rewritten as

$$\mathbb{E}\{R_{\mathrm{B}}^{(k)}\} - \mathbb{E}\{R_{\mathrm{E}}^{(k)}\} \geq \mathbb{E}\left\{\sum_{i=1}^{L_{\mathrm{u}}^{\max}} \log_2\left(\frac{p_t}{N \kappa_{\mathrm{B}}}\right) + \sum_{i=1}^{L_{\mathrm{u}}^{\max}} \log_2\left(|\mathbf{H}_{\mathrm{f}\,i,i}^{(k)}|^2\right)\right.$$
$$\left. + \sum_{i=L_{\mathrm{u}}^{\max}+1}^{N} \log_2\left(\frac{\frac{|\mathbf{H}_{\mathrm{f}\,i,i}^{(k)}|^2}{\kappa_{\mathrm{B}}}}{\frac{\sigma_{\mathrm{A-E}}^{(k)\,2}(L_{\mathrm{E}}^{(k)}+1)}{\kappa_{\mathrm{E}}}}\right) - \sum_{i=1}^{L_{\mathrm{u}}^{\max}} \log_2\left(\frac{1 + \frac{\frac{\alpha^{(k)}}{N} \sigma_{\mathrm{A-E}}^{(k)\,2}(L_{\mathrm{E}}^{(k)}+1)}{\beta_i\, K}}{\alpha^{(k)}}\right)\right\} \tag{74}$$

where $\frac{p_t}{N\kappa_{\mathrm{B}}}$ is the input SNR level per data symbol, which is assumed to be very high. Since the first term in (74) is very large and monotonically increasing with $p_t$ and all the other terms are very small, we can further ignore all terms except the first term. That is, the lower bound on the average secrecy rate (in bits/sec/Hz) is given by

$$\frac{1}{M+M_{\mathrm{cp}}}\mathbb{E}\left\{\left[R_{\mathrm{B}}^{(k)} - R_{\mathrm{E}}^{(k)}\right]^+\right\} \gtrapprox \frac{L_{\mathrm{u}}^{\max}}{M+M_{\mathrm{cp}}} \log_2\left(\frac{p_{\mathrm{t}}}{N \kappa_{\mathrm{B}}}\right) \tag{75}$$